\documentclass[reprint,%
 amssymb, amsmath, pre,
 aps,%cha%, p, numerical%
]{revtex4-1}

\usepackage{bm}%
\usepackage[colorlinks=true,linkcolor=blue]{hyperref}%
\usepackage{xcolor}

\usepackage{dcolumn}%
\usepackage{graphicx}
\expandafter\ifx\csname package@font\endcsname\relax\else
 \expandafter\expandafter
 \expandafter\usepackage
 \expandafter\expandafter
 \expandafter{\csname package@font\endcsname}%
\fi
\usepackage{subfigure}
\begin{document}

\title{Phase Diagram for a model of Spin-Crossover in Molecular Crystals}%
 
\author{J. Quetzalc\'oatl Toledo-Mar\'in}%
\email{j.toledo.mx@gmail.com}
\affiliation{Departamento de Sistemas Complejos, Instituto de
F\'{i}sica, Universidad Nacional Aut\'{o}noma de M\'{e}xico (UNAM),
Apartado Postal 20-364, 01000 M\'{e}xico, CDMX,
M\'{e}xico}%

\author{Carlos Rodriguez}%
\email{crc@fisica.uh.cu}
\affiliation{Faculty of Physics, University of Havana, 10400 Havana, Cuba}%
\affiliation{Centro de Investigacion en Ciencia Aplicada y Tecnologia Avanzada (CICATA-Legaria), Instituto Politecnico Nacional, 11500 Ciudad de Mexico, Mexico }

\author{Yosdel Plasencia Montesino}%
\email{laplace2108@gmail.com}
\affiliation{Centro de Investigacion en Ciencia Aplicada y Tecnologia Avanzada (CICATA-Legaria), Instituto Politecnico Nacional, 11500 Ciudad de Mexico, Mexico }

\author{Gerardo G. Naumis}%
\email{naumis@fisica.unam.mx}
\affiliation{Departamento de Sistemas Complejos, Instituto de
F\'{i}sica, Universidad Nacional Aut\'{o}noma de M\'{e}xico (UNAM),
Apartado Postal 20-364, 01000 M\'{e}xico, CDMX,
M\'{e}xico}%

\date{\today}
%\revised{?}

\begin{abstract}
Spin-crossover has a wide range of applications from memory devices to sensors. This has to do mainly with the nature of the transition, which may be abrupt, gradual or incomplete and may also present hysteresis. This transition alters the properties of a given sample, such as magnetic moment, color and electric resistance to name some. Yet, a thorough understanding of the phenomenon is still lacking. In this work  a simple model is provided to mimic some of the properties known to occur in spin-crossover. A detailed study of the model parameters is presented using a mean field approach and exhaustive Monte Carlo simulations. A good agreement is found between the analytical results and the simulations for certain regions in the parameter-space. This mean field approach breaks down in parameter regions where the correlations and cooperativity may no longer be averaged over.
\end{abstract}


\maketitle
%\tableofcontents

\section{Introduction}
Research around several phenomena in the overlap between solid state and condensed matter have the peculiar behavior of having periods of time where it is quiet and some other periods of time where a lot of research is being done. Such is the case of spin-crossover (SCO) phenomena, which spans broadly nine decades (see Ref. \cite{halcrow2013spin} for a compilation of research over the years together with Refs. \cite{gutlich2004spin, halcrow2011structure}). The SCO phenomenon is the the transition between a low spin
(LS) and a high spin (HS) state on a metal ion with $d^{4}-d^7$ electronic configuration. Experimental evidence suggest that the cause of this transition has to do with the competition between the strength of the field of
ligands and the spin-coupling energy between electrons \cite{shriver1994standard}. For instance, octahedral compounds Transition Metals Series $3d^{4-7}$ present this type of transition, in which they
can be in a high or low spin state, depending on whether the ligand field is
stronger or weaker than the matching energy. 

Furthermore, in the case of thermally induced SCO transition, the free energy difference between both states should be of the order of thermal energy, i.e., $k_B T$ (henceforth we consider $k_B=1$) \cite{gutlich1994thermal}. In this sense, high temperature favors HS whereas low temperature favors LS. Interestingly, the phenomenon is rather ubiquitous in nature. For instance, in the case of the transition in $\text{Fe}^{II}$ complexes between $t_{2g}^6$ ($S=0$) and $t_{2g}^6$ ($S=2$)
configurations, SCO is responsible for oxygen transport in hemoglobin and probably for the change under pressure of ferropericlase in the Earth's mantle \cite{yang2015elasticity}. In solids, SCO can be found in many transition metal oxides, organometallic complexes, inorganic salts or organic radicals and has a cooperative nature, frequently leading to abrupt changes of macroscopic physical properties and hysteresis. Perhaps even more appealing are the applications of SCO which range from display and memory devices and electroluminescent devices to MRI contrast agent \cite{letard2004towards,halcrow2013spin}. Additionally, the use of SCO combined with the properties of nanoporous metal-organic frameworks may also be used in molecular sensing \cite{halder2002guest}.

The HS and LS phases have different properties that depend on the electronic distribution in 3d orbitals. In this sense, the HS to LS transition has a large impact on the physical properties of a material, such as the magnetic moment, the color, the dielectric constant as well as the electrical resistance, among other properties. In other words, features such as optical, vibrational, magnetic and structural differ between one phase and the other. Hence, measuring these properties serve as a proxy to measure and monitor the SCO induced by external perturbation, such as light, pressure or temperature, for instance \cite{gutlich1990thermal, chernyshov2007coupling, konishi2008monte, ohkoshi2011light, pinkowicz2015enforcing}. Several techniques such as magnetic susceptibility measurements, as well as optical and vibrational spectroscopy of the kind of UV, IR, Raman and Mossbauer spectroscopy are used for this end. However, the holy grail is predicting the spin curve for a given material under cooling and heating together with the critical temperature and the hysteresis loop. There has been various efforts in this directions \cite{chernyshov2004landau,chernyshov2007coupling}, yet there is still a lack of a theory, which should not come as a surprise given the great variety of materials having SCO. A
complete microscopic description requires the consideration of three basic ingredients: i) the spin and vibrational states of individual octahedral complexes, ii) the interaction between them leading to cooperative effects and iii) the coupling with external factors such as temperature, pressure as well as external electric or magnetic  field . In this regard, for instance, in oxides cooperativity is
attributed to electronic exchange while in molecular crystals to electron-phonon coupling \cite{nesterov2017cooperative}. 
%Recent contributions [9] emphasize the key importance of strong coupling between eg electrons and molecular fully symmetric (“breathing”) vibration modes (instead of Jahn-Teller vibrations as previously proposed [10,11]) in determining the local energy pattern of SCO in molecular crystals.

In a previous paper by two of the authors \cite{rodriguez2018spin}, a theoretical approach to SCO in mononuclear molecular crystals containing $Fe^{II}$ ions was presented, where a simple effective interaction between neighboring local breathing modes was considered. Only electrons in eg states are linearly coupled to vibrations, being that the case there is no need for two breathing modes with different vibrational frequencies and coupling parameters. Furthermore, decoupling breathing modes with successive canonical transformations leads to a lattice model where short range and long range ferromagnetic and antiferromagnetic interactions arise in a strightforward manner. In the present paper we further study the phase diagram  by means of Monte Carlo simulations and analytical derivations. The structure of the paper is as follows: In the next section we present the model and its features, in section \ref{sec:MF} we solve the model in the thermodynamical limit, in section \ref{sec:Sim} we compare the analytics with the simulations and discuss the results, finally section \ref{sec:Conc} is for conclusions.

\section{Model}
Consider a $d$- dimensional periodic array of $i \; (1,...,N)$ mono-nuclear metal complexes, each containing an ion $\text{Fe}^{\text{II}}$ in an octahedral site surrounded by non-magnetic ligands. The number of electrons occupying $e_g$ states at the $i$th ion will be denoted with $n_i$ which represent the spin on the lattice site $i$. We further consider the degeneracy for each state, $g_n$, such that $g_0=1$, $g_1=9$ and $g_2=15$. Since the charge density distribution of the antibonding $e_g$ states is more localized near their octahedral neighbors, these electrons are strongly coupled with a local breathing vibration mode described by the operators $ \hat{a}_i^\dag$ and $ \hat{a}_i$. Breathing modes at neighboring sites interact via acoustic phonons. Thus, the effective Hamiltonian for this electron-local  vibrations system is:
\begin{equation}
\hat{H}=\sum_{i=1}^N \hat{H}_i + \sum_{\langle i,j \rangle} \hat{V}_{ij} \; ,
\end{equation}
where
\begin{equation}
\hat{H}_i=\epsilon n_i + \left( \hat{a}_i^\dag  \hat{a}_i + \frac{1}{2} \right) - \alpha n_i \left( \hat{a}_i^{\dag} +  \hat{a}_i \right) \; ,
\end{equation}
is the Hamiltonian for the $i$th ion, a harmonic oscillator and a term coupling both. Whereas,
\begin{equation}
\hat{V}_{ij} = -\frac{\lambda}{4} \left( \hat{a}_i^{\dag} +  \hat{a}_i \right)\left( \hat{a}_j^{\dag} +  \hat{a}_j \right)
\end{equation}
takes into account the coupling between breathing modes localized in neighbor sites. The parameter $\epsilon>0$ is the excitation energy per $e_g$ electron, which is obtained by subtracting the splitting energy between $t_{2g}$ and $e_g$ states and the pairing energy $P$ in $t_{2g}$ states, while $\alpha$ and $\gamma$ are coupling parameters. We will assume $\epsilon=10$ unless stated otherwise explicitly. Moreover, we consider the phonon energy of the local breathing mode when the ion is in HS-state $\hbar \omega=1$ as well as the Boltzmann constant $k_B=1$.

Coupling to local modes induces virtual transitions that renormalize electron energies, give rise to an effective electron-electron interaction and shift atomic positions (see below). The interaction Hamiltonian $\hat{V}_{ij}$ is the simplest approximation to an effective inter-site coupling
which could result from averaging over degrees of freedom (acoustical phonons) connecting local breathing modes at neighboring sites \cite{palii2015microscopic}.

Breathing modes can be decoupled by canonical transformations of their creation and annihilation operators,
leading to a system with an effective electron-electron interaction and independent dispersive phonons:
\begin{equation}
\hat{H}= \hat{H}_e+\hat{H}_{ph} \; ,
\end{equation}
where
\begin{eqnarray}
\hat{H}_e &=& \sum_{i=1}^N \left(\epsilon n_i - \alpha^2 n_i^2 \right) - \alpha^2 \lambda \sum_{\langle i,j \rangle} n_i n_j - \sum_{i,j} U_{ij} n_i n_j \; , \label{eq:HamElec} \\
\hat{H}_{ph}&=&\sum_{\bm{q}} \sqrt{|1-\lambda s(\bm{q})|} \left(\hat{a}_j^{\dag} \hat{a}_j + \frac{1}{2} \right) \; , \label{eq:HamElec1} \\
U_{ij}&=&\frac{\alpha^2 \lambda^2}{N} \sum_{\bm{q}} \frac{s^2(\bm{q})}{1-\lambda s(\bm{q})} e^{\imath \bm{q} \cdot \left(\bm{R}_i - \bm{R}_j \right)} \; , \label{eq:HamElec2} \\
s(\bm{q})&=&\frac{1}{2} \sum_{j(i)=1}^z e^{\imath \bm{q} \cdot \left(\bm{R}_i -\bm{R}_j \right)} \; . \label{eq:HamElec3}
\end{eqnarray}
Here, the notation $j(i)$ and $\langle i,j \rangle$ denote the first neighbor $j$ for a given $i$ and pairwise neighbors, respectively and $z$ is the coordination number, which we assume equal to $6$ whenever is not explicitly specified. The actual derivation of Eqs. \eqref{eq:HamElec}, \eqref{eq:HamElec1}, \eqref{eq:HamElec2} and \eqref{eq:HamElec3} is documented in SN 3.

\section{Mean field} \label{sec:MF}
In this section we present a mean field approximation of the electron-electron Hamiltonian produced in Eq. \eqref{eq:HamElec}. The idea is to consider first order fluctuations in spins, denoted as $\delta n_i$, while neglecting second order fluctuations as in Ref. \cite{cardy1996scaling}. To this end, we write $n_i=n-\delta n_i$ where $n$ is the mean value of the spins. It is easy to show that the electron-electron Hamiltonian in Eq. \eqref{eq:HamElec} becomes
\begin{eqnarray}
H_e^{mf} &=& \alpha^2 n^2 \left(1+\frac{\lambda z}{2} \right) + \sum_{i=1}^N \left(\epsilon-2n \alpha^2 \left(1+\frac{\lambda z}{2} \right) \right) n_i \nonumber \\
&& + n^2 \sum_{i,j}U_{ij} -2n \sum_i \sum_j U_{ij} n_i \; . \label{eq:HamMF}
\end{eqnarray}

Notice that under this approach, the spin-spin product no longer appears and, instead, spins interact with a mean field proportional to $\sim n$. Then, computing the partition function using the mean field Hamiltonian, using Eq. \eqref{eq:HamMF}, is direct, and yields: 
\begin{widetext}
\begin{eqnarray}
Z_{mf}&=& \exp \left[ -\beta \left(\alpha^2 \left(1+\frac{\lambda z}{2} \right)N + \sum_{i,j}U_{ij} \right) n^2 \right] \times \nonumber \\
&& \prod_{i=1}^N  \left(g_0+ g_1  e ^{-\beta \left(\epsilon -2 n \left( \alpha^2 \left(1+\frac{\lambda z}{2} \right) + \sum_{j}U_{ij} \right) \right) } + g_2  e ^{-2\beta \left(\epsilon -2 n \left( \alpha^2 \left(1+\frac{\lambda z}{2} \right) + \sum_{j}U_{ij} \right) \right) } \right) \label{eq:PartitionFunc}
\end{eqnarray}
\end{widetext}

Notice in the partition function (Eq. \eqref{eq:PartitionFunc}) how the dependence on the index $i$ from the product falls with the long range interaction, i.e., $U_{ij}$. In the case for $N \gg 1$ we may approximate the sum of long range interactions as
\begin{equation}
\sum_{i,j}U_{ij} \approx -N\frac{\alpha^2 \lambda z}{2} \frac{1}{1-\frac{2}{\lambda z}} \; . \label{eq:LRApprox}
\end{equation}

The derivation of Eq. \eqref{eq:LRApprox} is shown in the Supplementary Note (SN) 1.

Hence, using the approximation presented in Eq. \eqref{eq:LRApprox}  we further simplify the partition function to,
\begin{eqnarray}
&&Z_{mf}= \exp \left[ -\beta \frac{n^2 \alpha^2 N}{1-\frac{\lambda z}{2}} \right] \times \nonumber \\
 && \left(  g_0+g_1 e^{-\beta \left(\epsilon - \frac{2n \alpha^2}{1-\frac{\lambda z}{2}} \right)} + g_2 e^{-2\beta \left(\epsilon - \frac{2n \alpha^2}{1-\frac{\lambda z}{2}} \right)} \right)^N . \label{eq:PartitionFunc2}
\end{eqnarray}

From Eq. \eqref{eq:PartitionFunc2} we may compute any thermodynamic quantity straightforward. Although we are using a mean field approach, the mean field free energy will depend upon the mean spin, which is not a proper free energy. Nevertheless, it may be proved that in the thermodynamical limit ($N \rightarrow \infty$) the minimum mean field free energy value as a function of $n$ equals the actual free energy \cite{kardar2007statistical}.

The mean field free energy per molecule, which we denote as $f_{mf}(n,T)$, is
\begin{eqnarray}
f_{mf}(n,T)&=&\frac{n^2 \alpha^2}{1-\frac{\lambda z}{2}} - T \ln \left( g_0+g_1 e^{-\beta \left(\epsilon - \frac{2n \alpha^2}{1-\lambda z/2} \right)} \right. \nonumber \\
&& \left.+ g_2 e^{-2\beta \left(\epsilon - \frac{2n \alpha^2}{1-\lambda z/2} \right)} \right) \; . \label{eq:FreeEnergyMF}
\end{eqnarray}

Similarly, in the case where we neglect long-range interaction, it is not difficult to realize that the mean field free energy, $f_{mf}^{SR}(n,T)$, yields:
\begin{eqnarray}
f_{mf}^{SR}(n,T)&=&n^2 \alpha^2 \left(1+\frac{\lambda z}{2} \right) \nonumber \\
&&- T \ln \left( g_0+g_1 e^{-\beta \left(\epsilon - 2n \alpha^2 \left(1+\frac{\lambda z}{2} \right) \right)} \right. \nonumber \\
&&  \left.+ g_2 e^{-2\beta \left(\epsilon - 2n \alpha^2 \left(1+\frac{\lambda z}{2} \right) \right)} \right) \; . \label{eq:FreeEnergyMF-SR}
\end{eqnarray}

Notice from Eqs. \eqref{eq:FreeEnergyMF} and \eqref{eq:FreeEnergyMF-SR} that for $\lambda \ll 2/z$, it follows $f_{mf} \approx f_{mf}^{SR}$. Conversely, under this approximation when $\lambda > 2/z$ the long range interaction does not decay with distance. Hence, this imposes a constraint over the system which makes it impossible to draw conclusions from a mean field approach and, in fact, we shall see that the system is highly size-dependent (this is shown and discussed in SN 1).

\begin{figure}[hbtp]
\centering
\subfigure[]{
\includegraphics[width=3.1in]{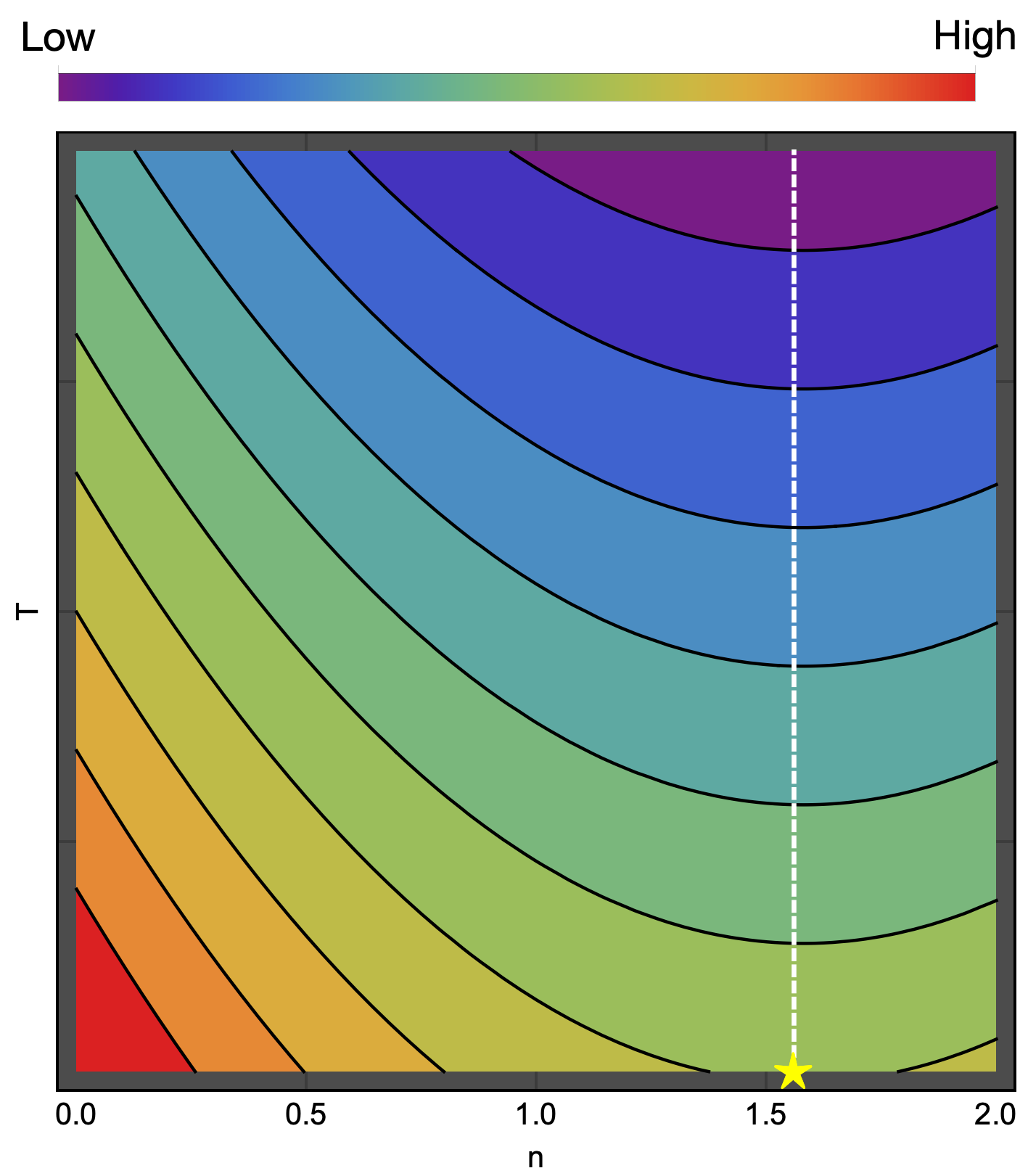}
\label{fig:FEMFLRInt-HighT-SmallL}}
\subfigure[]{
\includegraphics[width=3.1in]{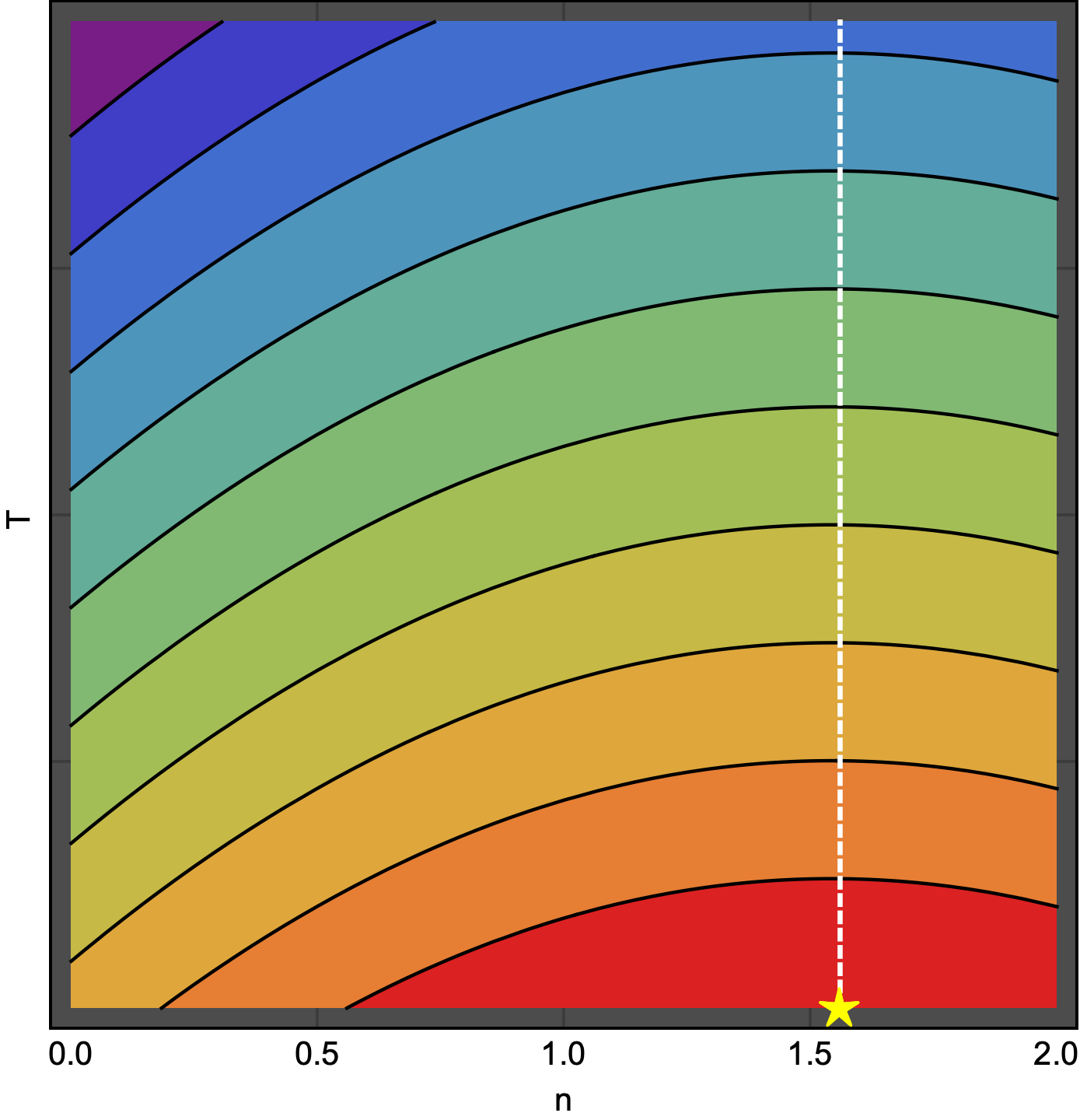}
\label{fig:FEMFLRInt-HighT-LargeL}}
\caption{Mean field free energy per molecule considering long range interactions given by Eq. \eqref{eq:FreeEnergyMF}. We have fixed the values $z=6, \; \epsilon=10, \; \alpha= \sqrt{5/2}+1/2$ with \textbf{(a)} $ \lambda=0.25 \; (<2/z)$ and \textbf{(b)} $ \lambda=0.5 \; (>2/z)$ while the temperature $T>700$. The yellow star corresponds to $n=\eta$, while the white line is for guide purposes. For high temperatures, when $\lambda<2/z$ the mean spin is $\eta$, whereas for $\lambda > 2/z$ the mean spin is $0$ (see text for discussion).} \label{fig:FEMF-HighT}
\end{figure}

Now, in the case of high temperatures, i.e., $T \gg \epsilon - 2n \alpha^2/ \left(1-\lambda z/2 \right)$, the mean field free energy becomes
\begin{equation}
f_{mf} \approx \epsilon \eta - \frac{ \alpha^2}{1-\frac{\lambda z}{2}} \left( \eta^2 - \left(n-\eta \right)^2 \right) - T \ln \sum_{i=0}^2 g_i \; ,\label{eq:FreeEnergyMFHighTemp}
\end{equation}
with $\eta=\sum_{i=0}^2 g_i n_i/\sum_{i=0}^2 g_i$ being the high temperature mean spin. Notice that the previous Eq. \eqref{eq:FreeEnergyMFHighTemp} has a minimum at $n=\eta$ provided $\lambda<2/z$. This is shown in Fig. \ref{fig:FEMFLRInt-HighT-SmallL} where we present a contour plot of the mean field free Energy per particle (Eq. \eqref{eq:FreeEnergyMF}) in a $T-n$ diagram. Moreover, this is in agreement with the fact that for small values of $\lambda$, $f_{mf} \approx f_{mf}^{SR}$ and in the case of short range interactions, at high temperatures the system must behave as a set of uncoupled spins. This implies that the mean spin is given by $\sum_{i=0}^2 p_i n_i$, where $p_i=g_i/\sum_{m=0}^2 g_m$ is the probability obtained simply from the degeneracy of each configuration neglecting any coupling in the model.

An interesting feature of the model is the fact that in the case where $\lambda>2/z$, the point $n=\eta$ becomes a maximum and the global minimum is located at $n=0$, i.e., at high temperatures the mean spin becomes zero. This is shown in Fig. \ref{fig:FEMFLRInt-HighT-LargeL} where we present a contour plot of the mean field free Energy per particle (Eq. \eqref{eq:FreeEnergyMF}) in a $T-n$ diagram. Thus, the condition $\lambda > 2/z$ preempts the maximization of entropy at high temperatures. In the SN 1 we show that, in fact, for $\lambda>2/z$ the long range interactions go as $N^{1/4} \exp(N)$, hence any mean field approach fails in describing the model in this parameter region. Hence, we should constrict the model to $\lambda < 2/z$.

\subsection*{Critical Temperature}
In this section we compute the critical temperature. In particular, we show that there are some parameter values in which at low temperatures the system is equally likely to have a transition where all spins are $2$ or $0$ mimicking a two-level system. To understand this, let us consider the mean field free energy in the case of short range interactions (Eq. \eqref{eq:FreeEnergyMF-SR}), however, the procedure is the same in the case of long range interactions mean field free energy. Notice that the exponentials inside the logarithm may go to zero or infinity for low temperatures depending on the values of the parameters as well as the mean spin. Let us consider low temperatures and that $n < \epsilon/2 \alpha^2 \left(1+\lambda z/2 \right)$. Then, for low temperatures, we may approximate Eq. \eqref{eq:FreeEnergyMF-SR} with solely the first term. Similarly, in the case where $n >\epsilon/2 \alpha^2 \left(1+\lambda z/2 \right) $ we may approximate Eq. \eqref{eq:FreeEnergyMF-SR} thus to write:
\begin{equation}
f_{mf}^{SR} \approx 
\begin{cases}
n^2 \alpha^2 \left(1+\frac{\lambda z}{2} \right) \; , \qquad \text{for } n < \epsilon/2 \alpha^2 \left(1+\lambda z/2 \right) \\
(n-2)^2 \alpha^2 \left(1+\frac{\lambda z}{2} \right) +2 \epsilon - T \ln g_2 -4 \alpha^2 \left(1+\frac{\lambda z}{2} \right) \; , \\
 \qquad \text{for } n > \epsilon/2 \alpha^2 \left(1+\lambda z/2 \right)
\end{cases} \label{eq:FreeEnergyMF-SR-Approx}
\end{equation}

\begin{figure}[hbtp]
\centering
\subfigure[]{
\includegraphics[width=3.2in]{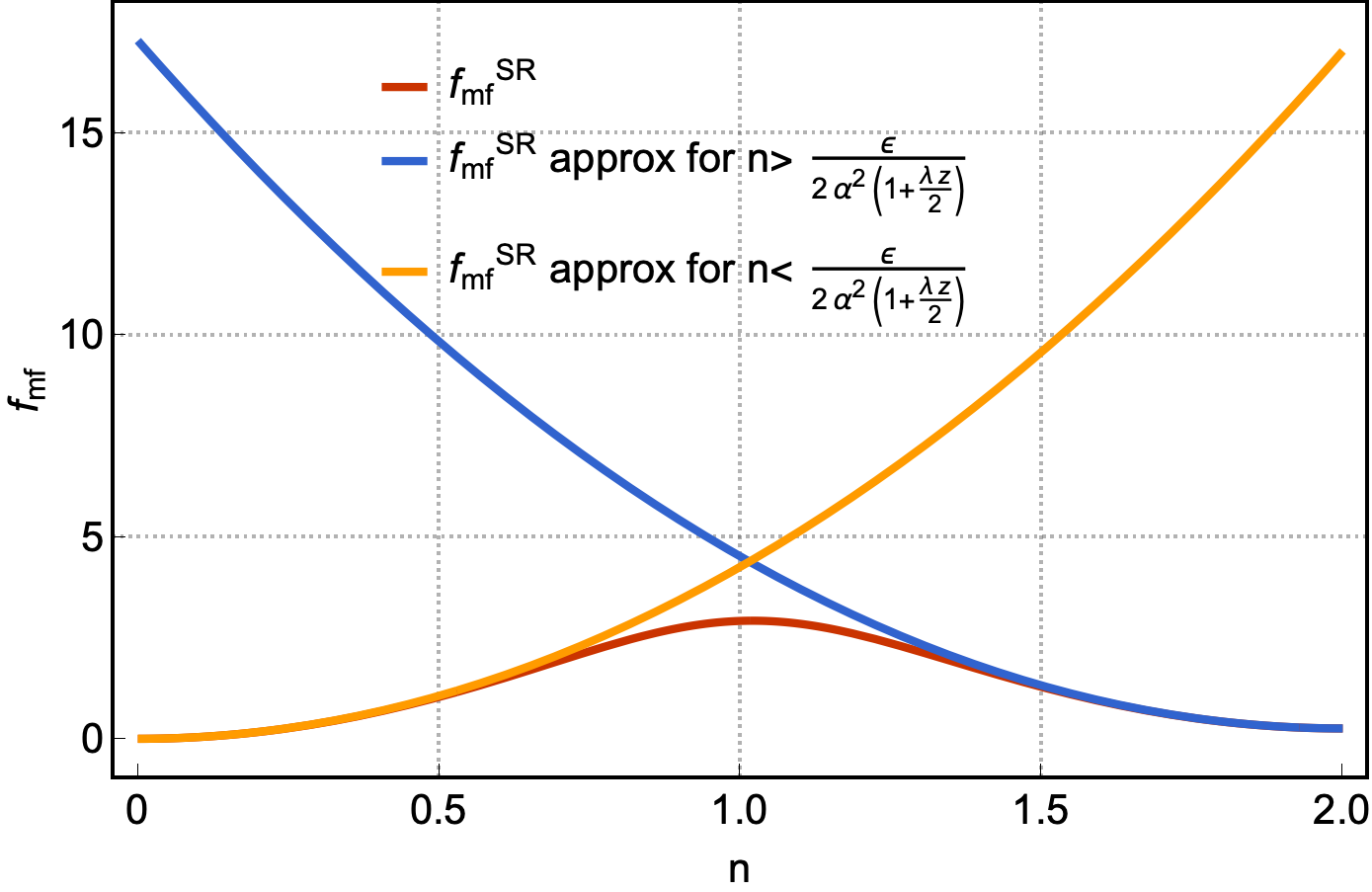}
\label{fig:CT-a}}
\subfigure[]{
\includegraphics[width=3.2in]{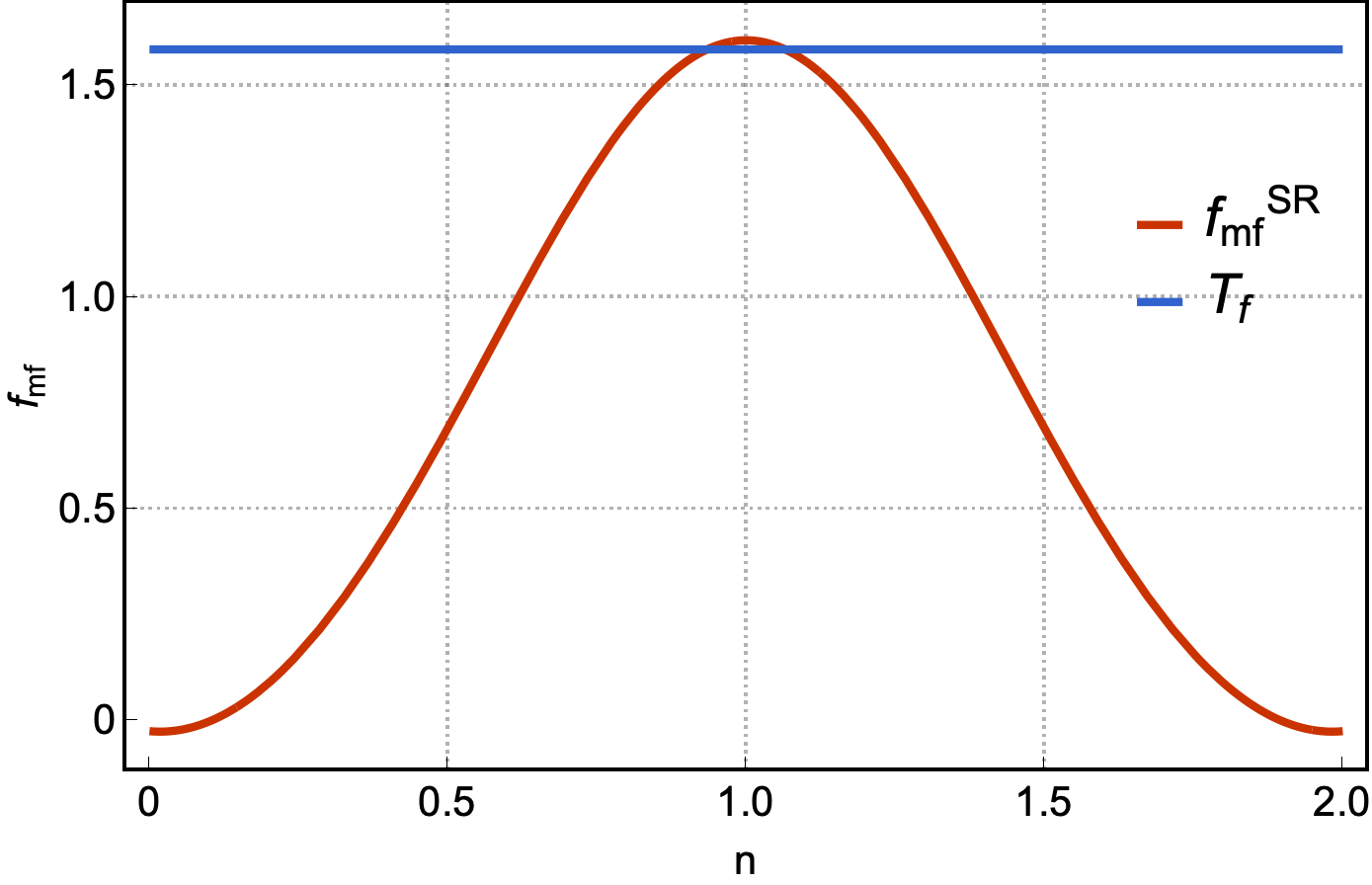}
\label{fig:CT-b}}\\
\caption{ \textit{(a)} Comparison between the mean field free energy (Eq. \eqref{eq:FreeEnergyMF-SR}) and the approximation at low temperatures (Eq. \eqref{eq:FreeEnergyMF-SR-Approx}) \textit{vs} n. We have fixed the values $z=6, \; \epsilon=10, \; T=1, \; \alpha=1.56$ and $\lambda=0.25$. \textit{(b)} Mean field free energy (Eq. \eqref{eq:FreeEnergyMF-SR}) evaluated at the temperature $T_f$ which guarantees that low spin and high spin have the same free energy. The plot also shows $T_f$ for a fixed parameter set $z=6, \; \epsilon=10, \; \alpha=1.56$ and $\lambda=0.25$. Notice that $T_f \sim f_{mf}^{SR}(T_f)$, which allows the system to choose between low spin or high spin.} \label{fig:CT}
\end{figure}

In Fig. \ref{fig:CT-a} we show the comparison between Eq. \eqref{eq:FreeEnergyMF-SR} and the approximation \eqref{eq:FreeEnergyMF-SR-Approx} for some fixed parameter values. Now, given we are using mean field, the minima of the mean field free energy should correspond with the actual free energy. Thus, we should expect that the system is likely to choose low or high spin for the same given parameter-set when two things are fulfilled, namely,
\begin{itemize}
\item The mean field free energy at low spin is equal to the mean field free energy at high spin.

\item The temperature is of the order of the maximum of the free energy.
\end{itemize}
The first condition translates into equating Eq. \eqref{eq:FreeEnergyMF-SR-Approx} for $n=0$ with itself for $n=2$. From this we obtain the temperature $T_f=2 \left(\epsilon-2 \alpha^2 \left(1+\lambda z/2 \right) \right)/\ln g_2$. The second condition implies that $T_f$ should be of the order of the maximum value in the free energy. In Fig. \ref{fig:CT-b} we have plotted the mean field free energy (Eq. \eqref{eq:FreeEnergyMF-SR-Approx}) at temperature $T=T_f$ as well as the temperature $T_f$ for a given set of parameters (see caption), in particular, we fixed $\alpha \approx 1.5$. Notice that the maximum of the free energy is of the order of the temperature which is $\approx 1.5$. As we will see later, the numerical simulations yield a critical temperature of $\approx 1$ for the same parameter values.

\begin{figure}[hbtp]
\centering
\centering
\subfigure[]{
\includegraphics[width=3.2in]{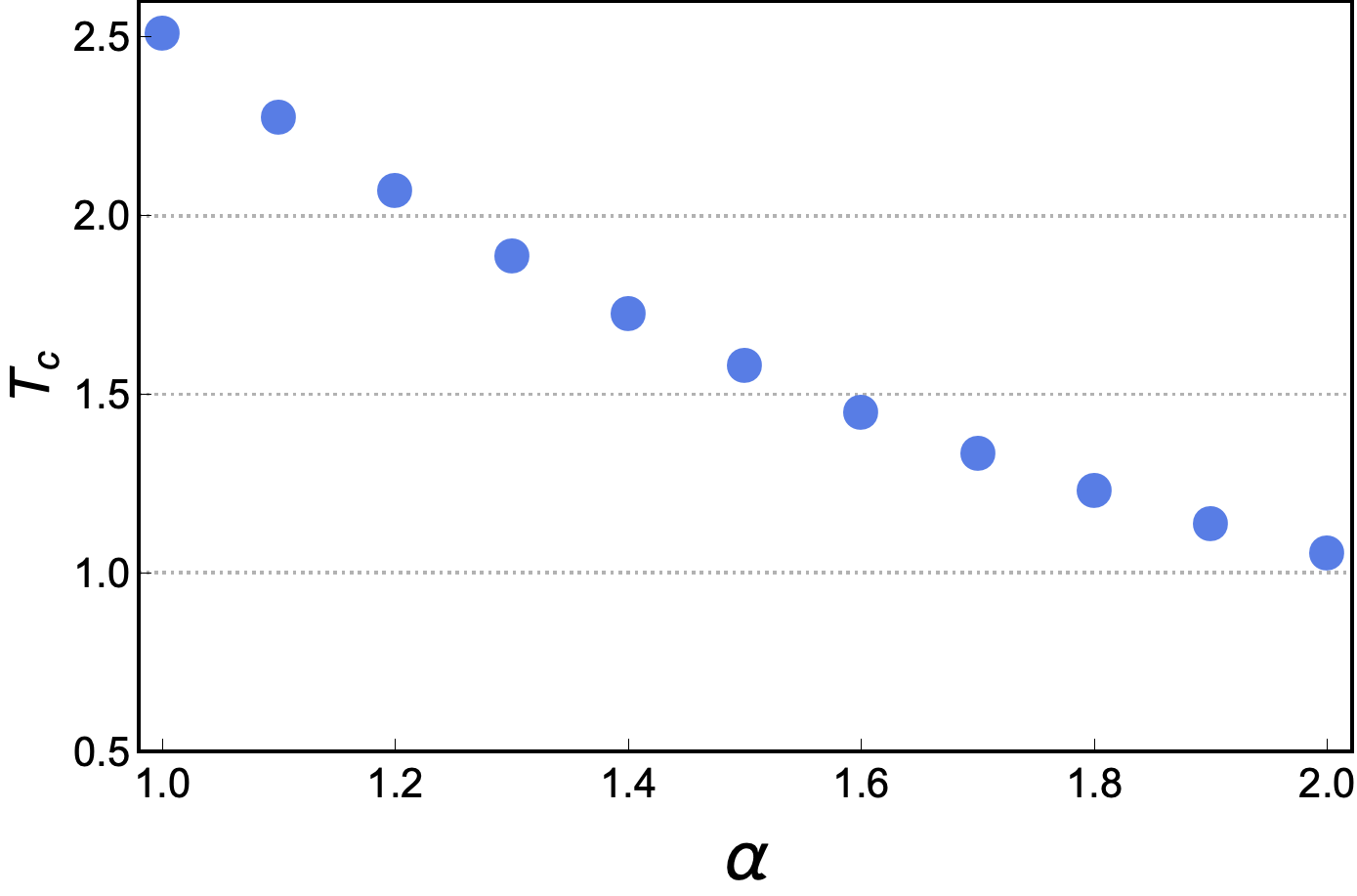}
\label{fig:Tc}}
\centering
\subfigure[]{
\includegraphics[width=3.2in]{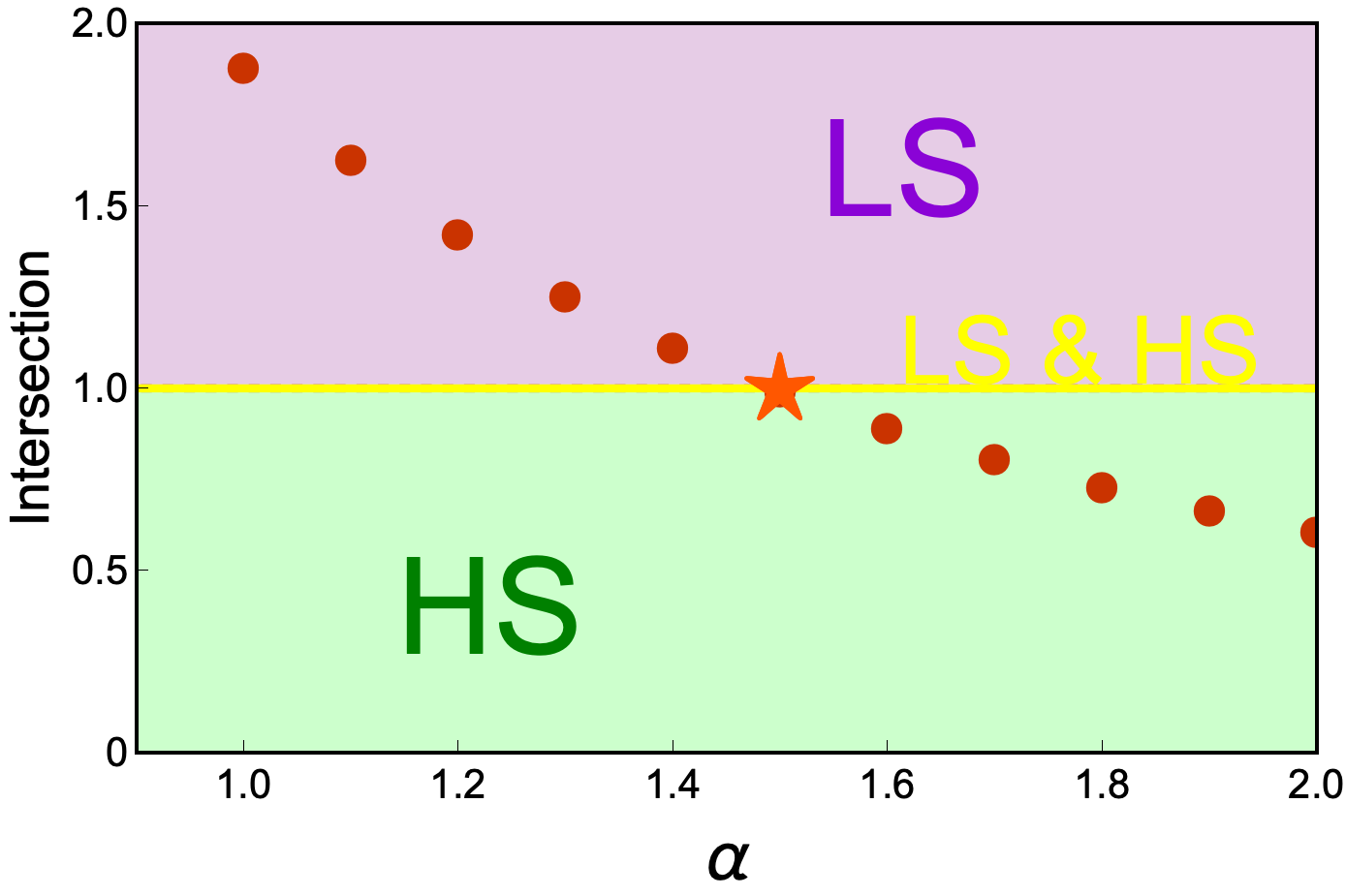}
\label{fig:Inter}}
\caption{\textbf{(a)} Plot of the critical temperature $T_c$ \textit{vs} $\alpha$ obtained as described in the text in the mean field approximation neglecting the long interactions (Eq. \eqref{eq:FreeEnergyMF}). \textbf{(b)} Plot of $n_{int}(T_c)$ (Eq. \eqref{eq:nint}) \textit{vs} $\alpha$. When $n_{int}(T_c)>1$ the system goes to LS when the transition occurs. On the contrary, when $n_{int}(T_c)<1$ the system goes to HS configuration when the transition occurs. When $n_{int}=1$ the system may go to LS or HS (see main text). The rest of the parameters were kept fixed at $\epsilon=10, \; z=6$ and $\lambda=0.25$.}
\end{figure}

We may further compute the critical temperature under this criteria. We do this as follows: first, we locate the intersection between Eq. \eqref{eq:FreeEnergyMF-SR-Approx} when $n<  \epsilon/2 \alpha^2 \left(1+\lambda z/2 \right)$ and itself when  $n>  \epsilon/2 \alpha^2 \left(1+\lambda z/2 \right)$. Let us denote this intersection as $n_{int}(T)$. It is a feasible task to obtain
\begin{equation}
n_{int}(T)=\frac{2\epsilon -T \ln g_2}{4 \alpha^2 \left(1+\frac{\lambda z}{2} \right)} \; . \label{eq:nint}
\end{equation}
Then, solving for $T$ the Eq. 
\begin{equation}
T=f_{mf}^{SR}(n_{int}(T),T)\; ,
\end{equation}
yields the critical temperature $T_c$. In Fig. \ref{fig:Tc} we have plotted the critical temperature \textit{vs} $\alpha$ from which one may appreciate that $T_c$ decreases as $\alpha$ increases, which is qualitatively consistent with the numerical simulations to be discussed in the next section. 

Additionally, we may predict if the transition is to LS or HS. This is done by noticing that if the intersection  between Eq. \eqref{eq:FreeEnergyMF-SR-Approx} when $n<  \epsilon/2 \alpha^2 \left(1+\lambda z/2 \right)$ and itself when  $n>  \epsilon/2 \alpha^2 \left(1+\lambda z/2 \right)$ happens at $n > 1$ then the mean field free energy has a minimum at $n=0$. Conversely, if it happens at $n<1$ then the minimum occurs at $n=2$. In Fig. \ref{fig:Inter} we have plotted $n_{int}(T_c)$ \textit{vs} $\alpha$. Notice that for $\alpha < 1.5$ the system goes to LS while for $\alpha > 1.5$ the system goes to HS. However, for $\alpha \approx 1.5$ the system is likely to go HS or LS.

A similar analysis may be done in case of $\lambda <2/z$ with long range interactions and the outcome is essentially the same. However, the critical temperature is somewhat lower than when neglecting long range interactions.

\begin{table}[!ht]
  \begin{center}
    \caption{Mean spin at low temperature.}
    \label{tab:table1}
    \subfigure[\; Considering long range interactions.]{
    \begin{tabular}{c|c|c} % <-- Alignments: 1st column left, 2nd middle and 3rd right, with vertical lines in between
     % \textbf{Value 1} & \textbf{Value 2} & \textbf{Value 3}\\
      $\alpha$ & $\lambda$ & $n$ \\
      \hline
      $\sqrt{5/2}$ & $0.25$ & $2$\\
      $\sqrt{5/2}+1/2$ & $0.25$ & $2$\\
      $\sqrt{5/2}-1/2$ & $0.25$ & $0$ or $2$\\
       $\sqrt{5/2}$ & $0.5$ & $2$\\
      $\sqrt{5/2}+1/2$ & $0.5$ & $2$\\
      $\sqrt{5/2}-1/2$ & $0.5$ & $2$\\
    \end{tabular}} 
    \subfigure[\; Without long range interactions.]{
     \begin{tabular}{c|c|c} % <-- Alignments: 1st column left, 2nd middle and 3rd right, with vertical lines in between
     % \textbf{Value 1} & \textbf{Value 2} & \textbf{Value 3}\\
      $\alpha$ & $\lambda$ & $n$ \\
      \hline
      $\sqrt{5/2}$ & $0.25$ &$0$ or $2$\\
      $\sqrt{5/2}+1/2$ & $0.25$ & $2$\\
      $\sqrt{5/2}-1/2$ & $0.25$ & $0$\\
    \end{tabular}
    }
  \end{center}
\end{table}

%\begin{table}[!ht]
  %\begin{center}
%    \caption{Spin at low temperature }
%    \label{tab:table1}
 %   \begin{tabular}{c|c|c} % <-- Alignments: 1st column left, 2nd middle and 3rd right, with vertical lines in between
     % \textbf{Value 1} & \textbf{Value 2} & \textbf{Value 3}\\
%      $\alpha$ & $\lambda$ & $n$ \\
   %   \hline
    %  $\sqrt{5/2}$ & $0.25$ & $2$\\
     % $\sqrt{5/2}+1/2$ & $0.25$ & $2$\\
     % $\sqrt{5/2}-1/2$ & $0.25$ & $0$ or $2$\\
%    \end{tabular}
%  \end{center}
%\end{table}

\section{Discussion} \label{sec:Sim}
In this section we compare the mean field results with the numerical simulations. This were done in C++ and Julia using a Monte Carlo algorithm with a Metropolis test \cite{Newman99} for system sizes $10^3, \; 15^3, \; 20^3$ and $100^3$ spins with periodic boundary conditions. We fixed the parameters $\epsilon=10$ and $z=6$ while considering, both, $\lambda=0.25 (<2/z)$ and $\lambda=0.5 (>z/6)$ and then for a given fixed set of the previous values we fixed $\alpha$ in the range from $0.1$ to $3.0$ in the simulations performed in Julia, while in the simulations performed in C++ we increased that range to $4.0$. We initiate the simulations at $T_{in}=15$ with all spins having a value equal to $2$. Then, we start lowering the temperature in steps of $\Delta T = 0.25$ and $\Delta T=1.0$ in the numerical simulation done in C++ and Julia, respectively, until reaching $T\approx 0$. 

\begin{widetext}

\begin{figure}[hbtp]
\centering
\subfigure[]{
\includegraphics[width=2.1in]{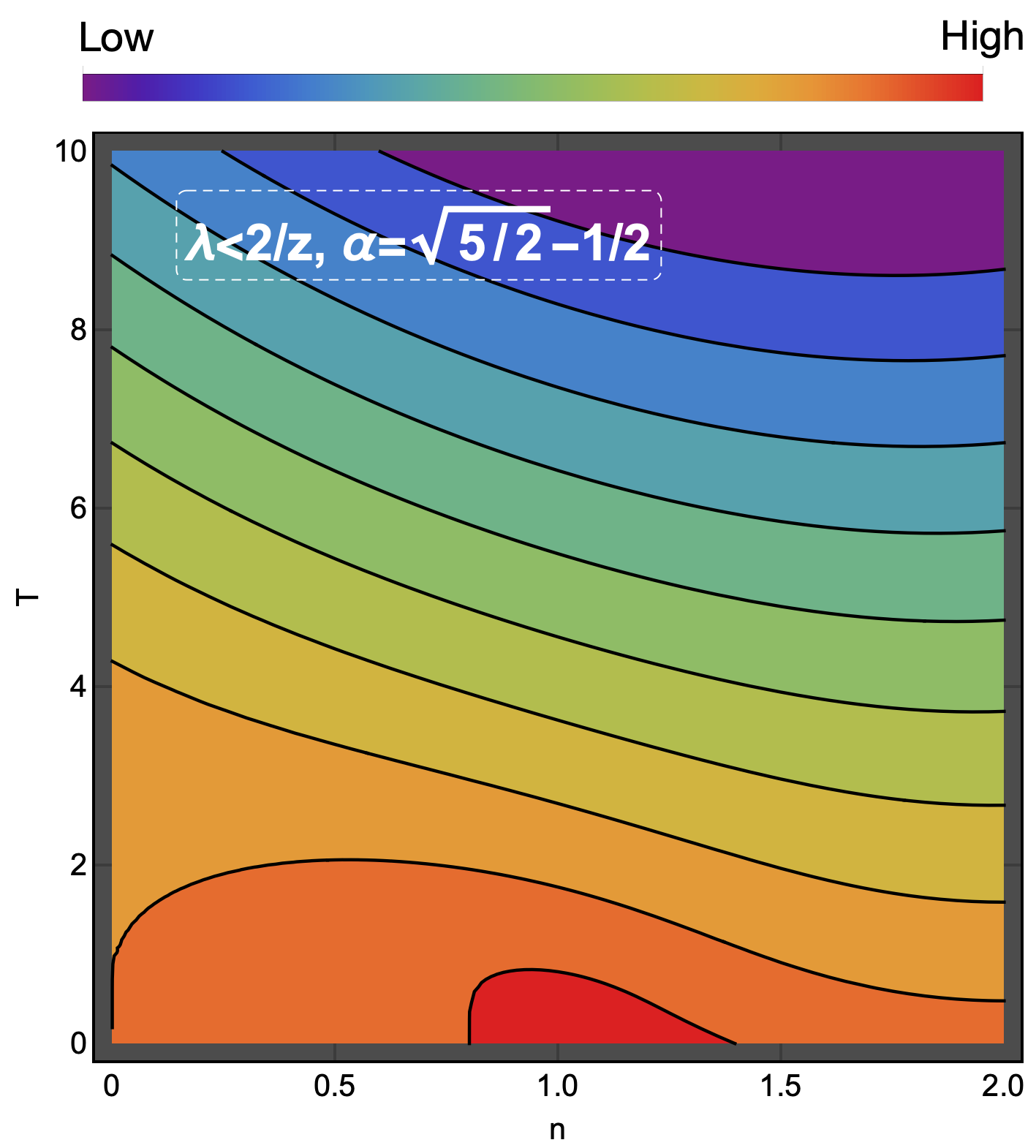}
\label{fig:FEMFLRInt-1}}
\subfigure[]{
\includegraphics[width=2.1in]{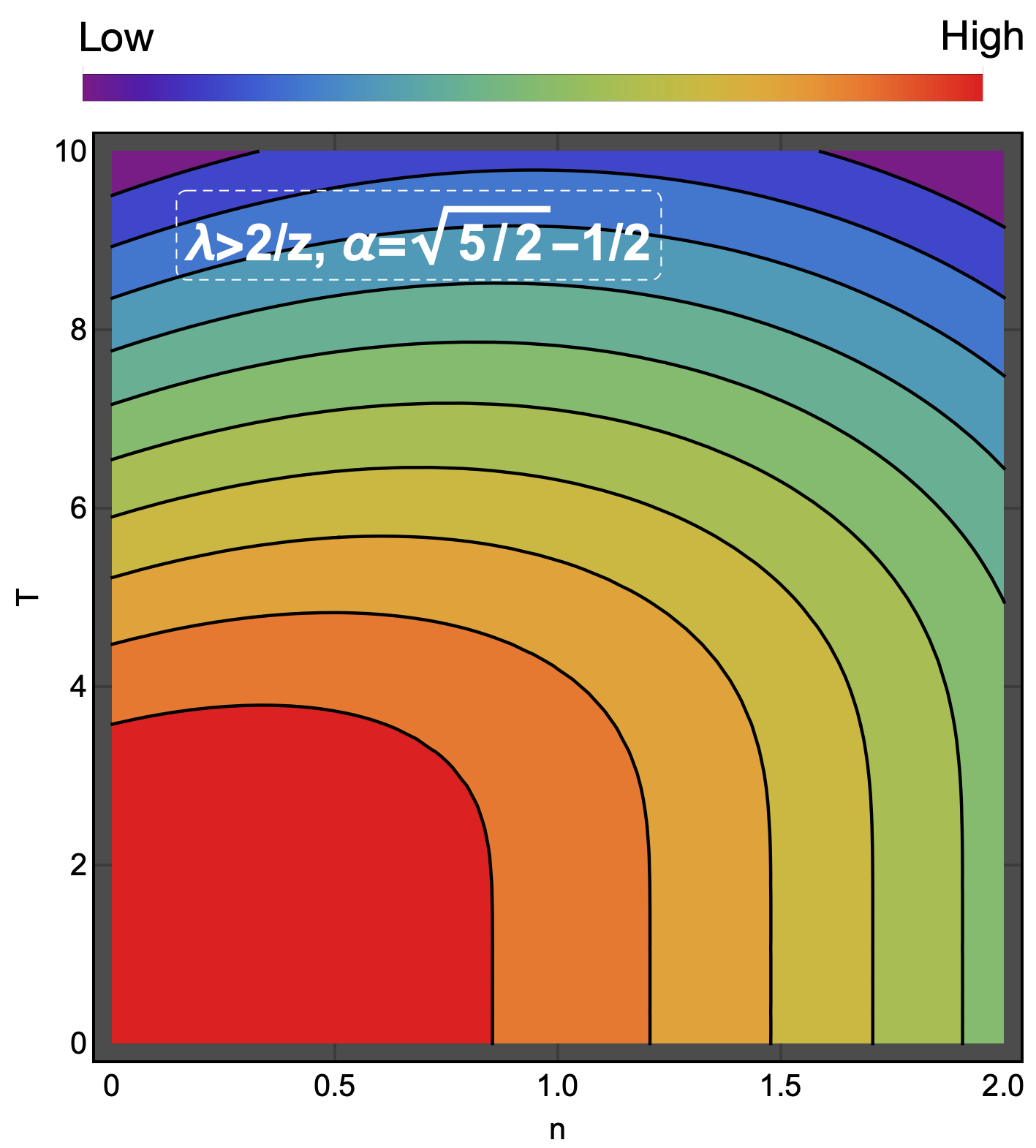}
\label{fig:FEMFLRInt-2}}
\subfigure[]{
\includegraphics[width=2.1in]{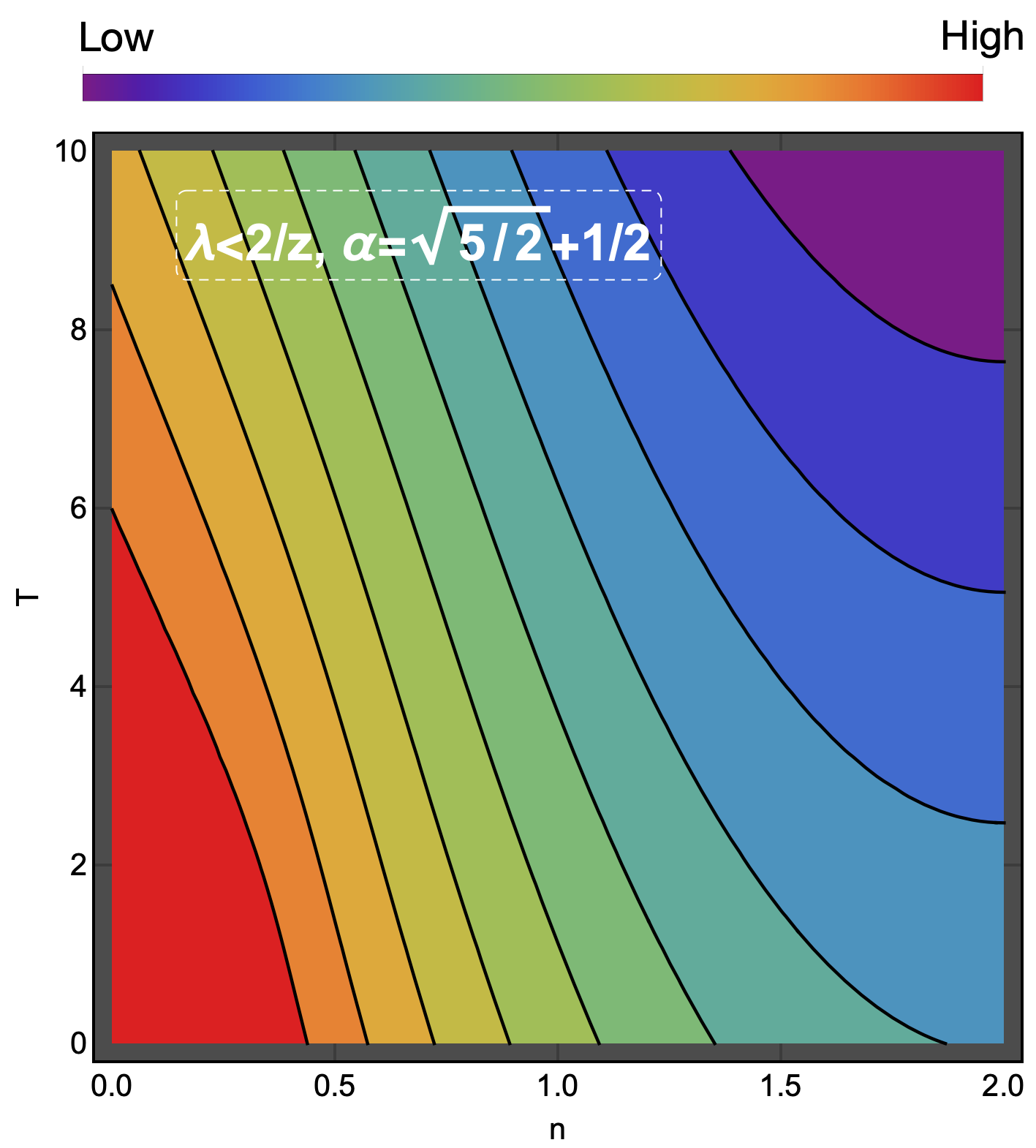}
\label{fig:FEMFLRInt-3}}
\subfigure[]{
\includegraphics[width=2.1in]{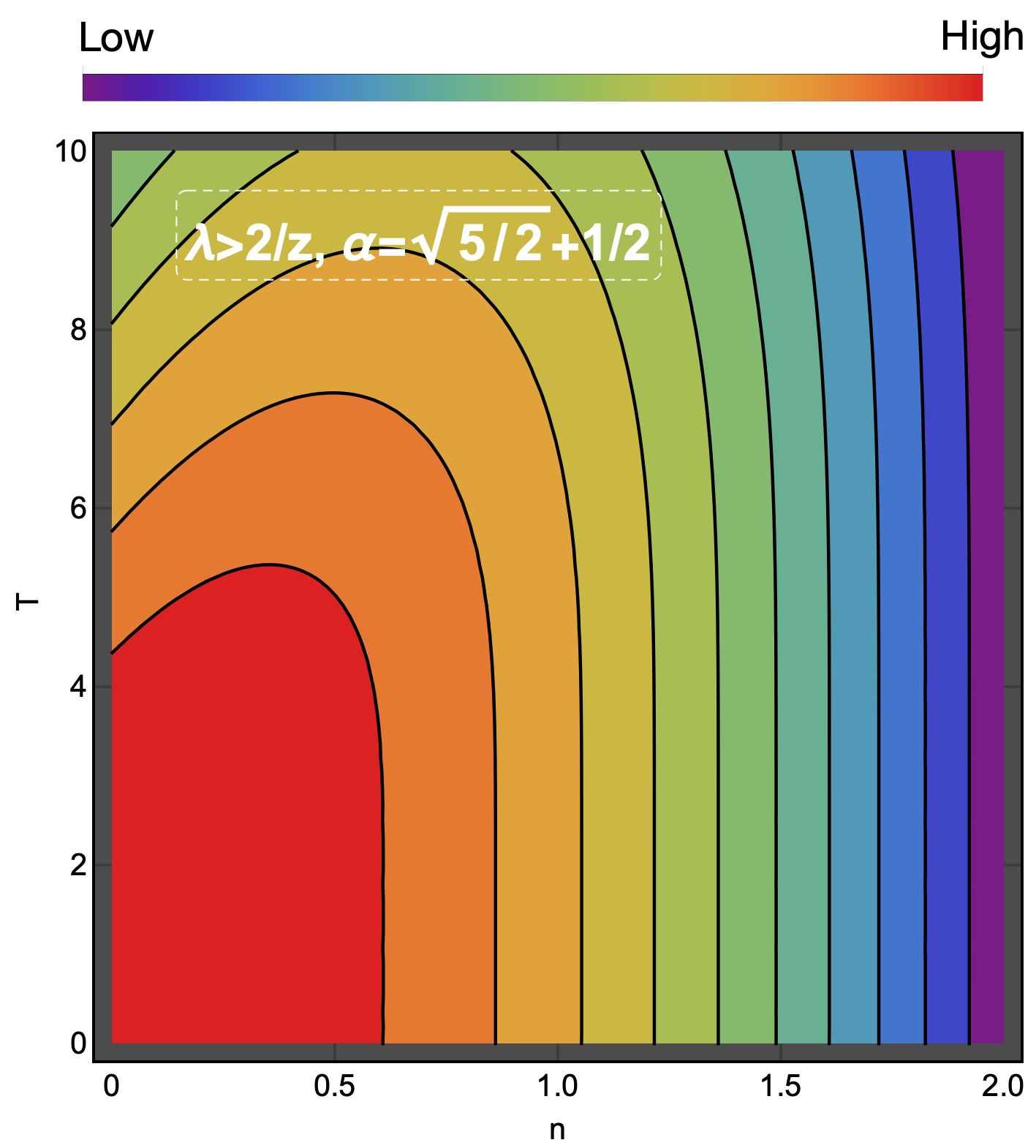}
\label{fig:FEMFLRInt-4}}
\subfigure[]{
\includegraphics[width=2.1in]{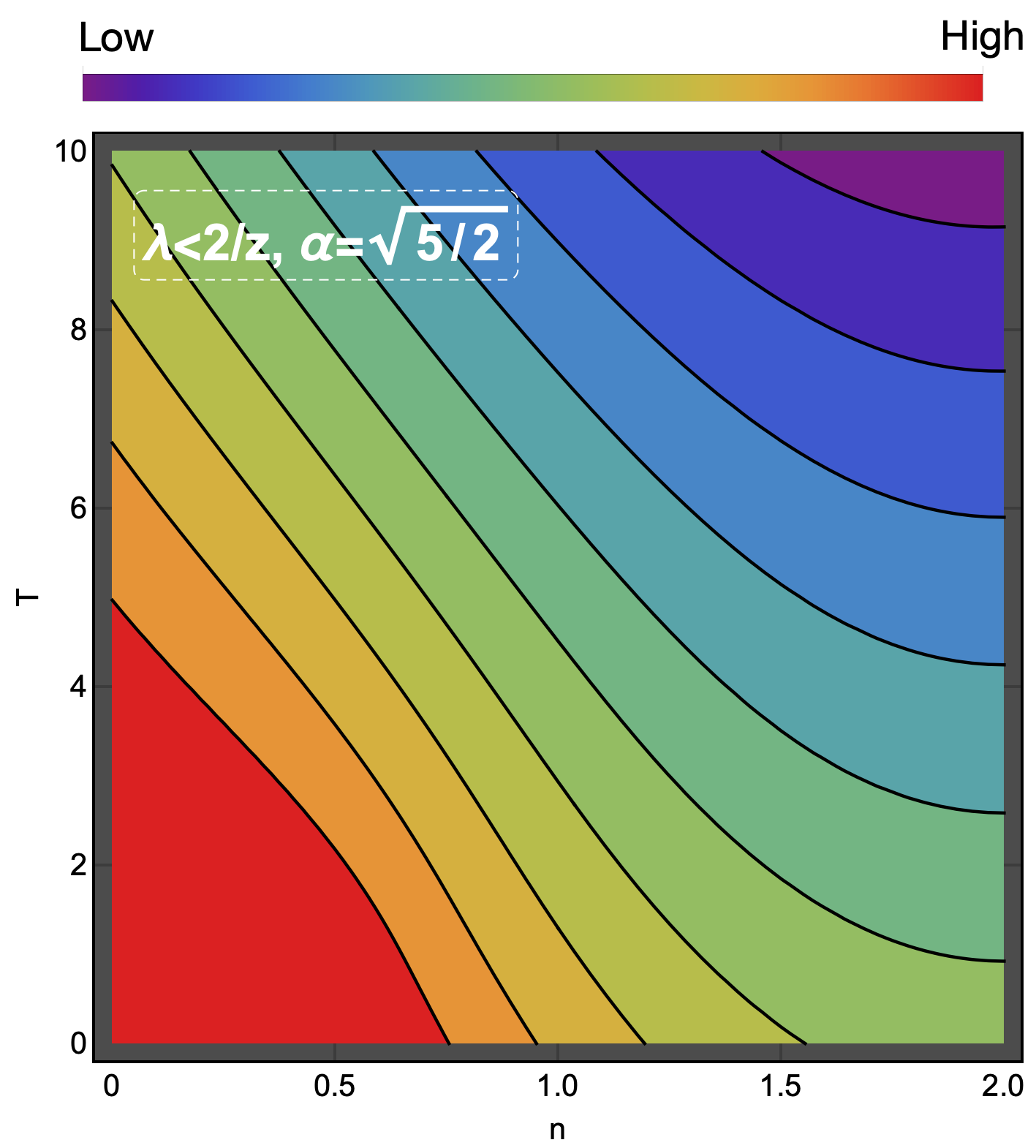}
\label{fig:FEMFLRInt-5}}
\subfigure[]{
\includegraphics[width=2.1in]{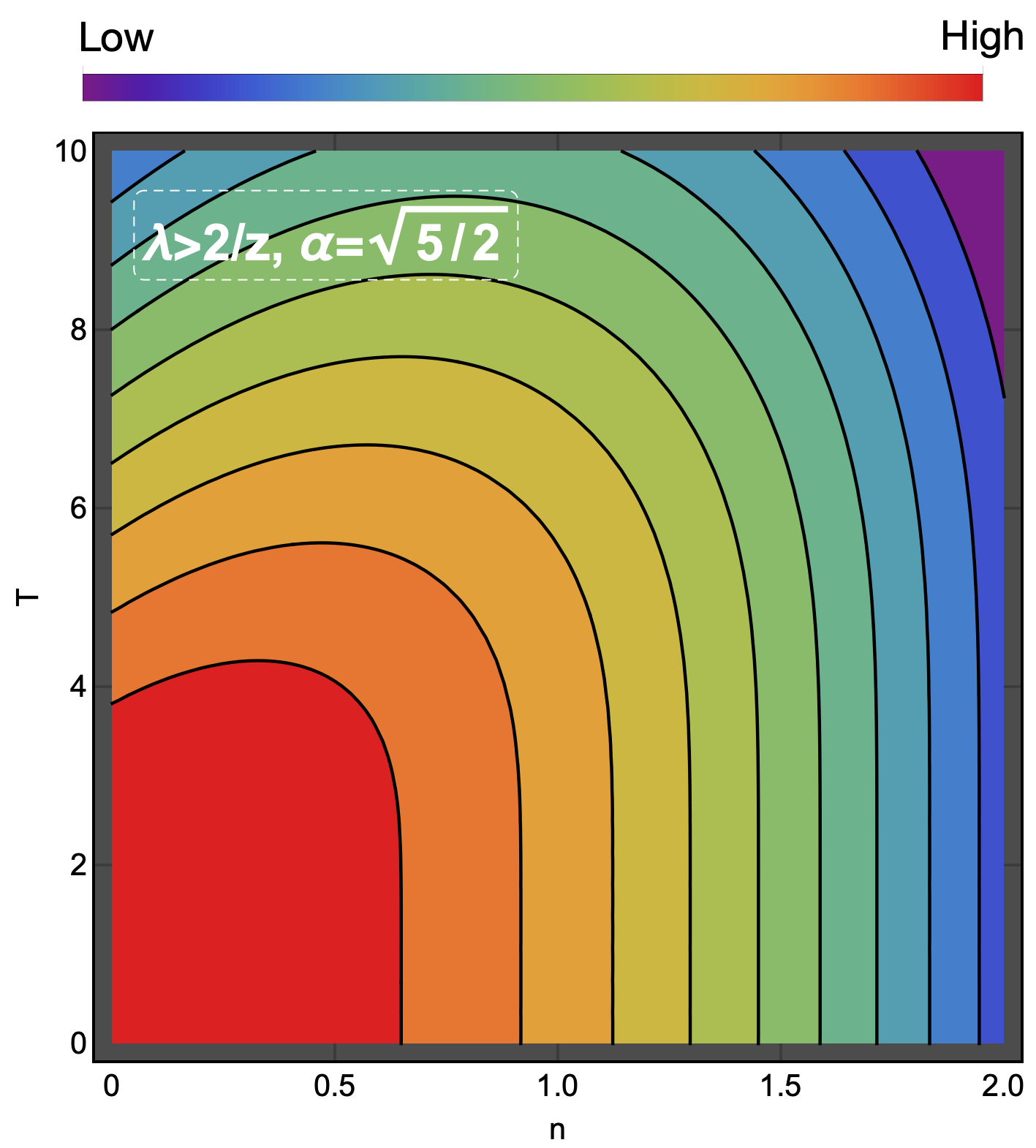}
\label{fig:FEMFLRInt-6}}
\subfigure[]{
\includegraphics[width=2.1in]{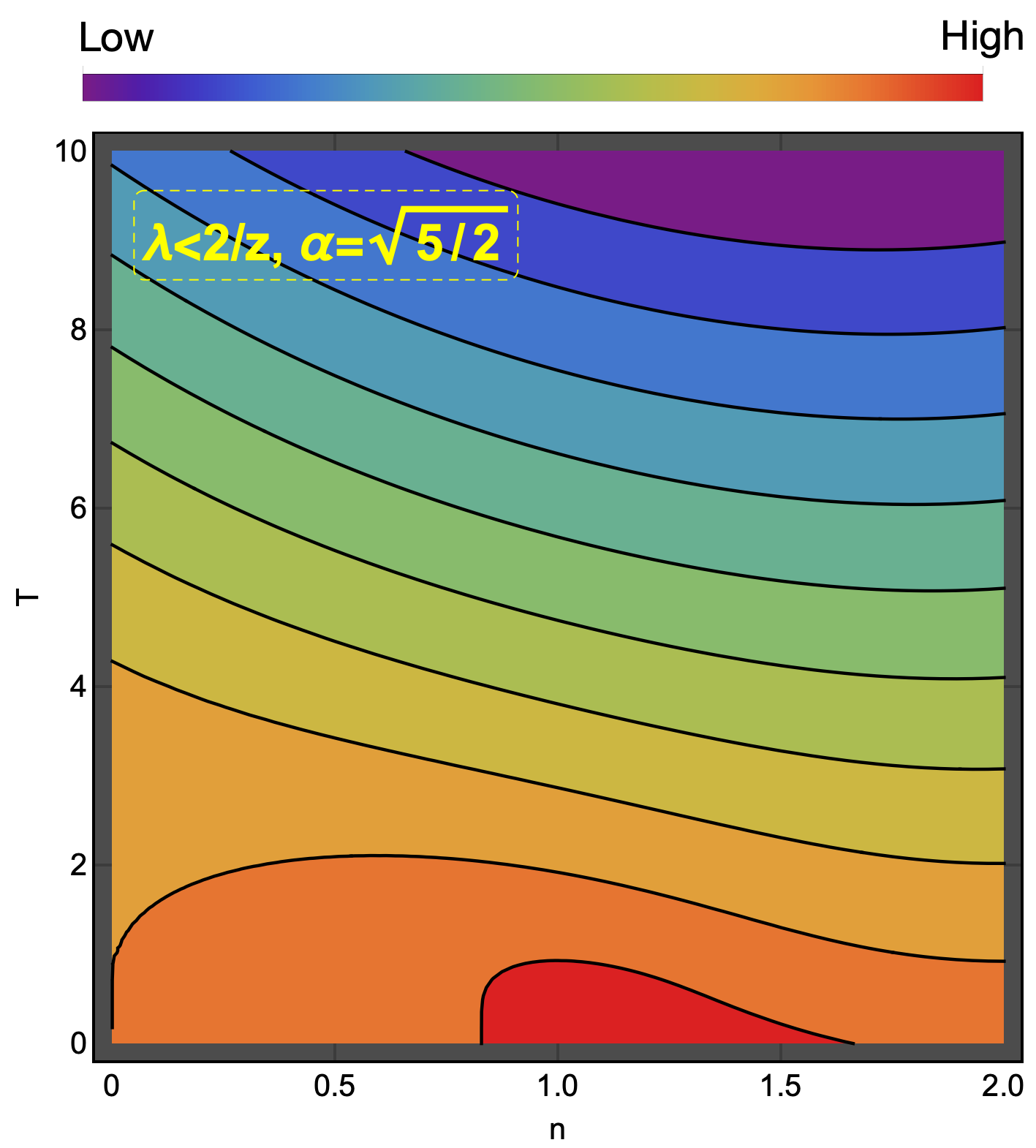}
\label{fig:FEMFLRInt-7}}
\subfigure[]{
\includegraphics[width=2.1in]{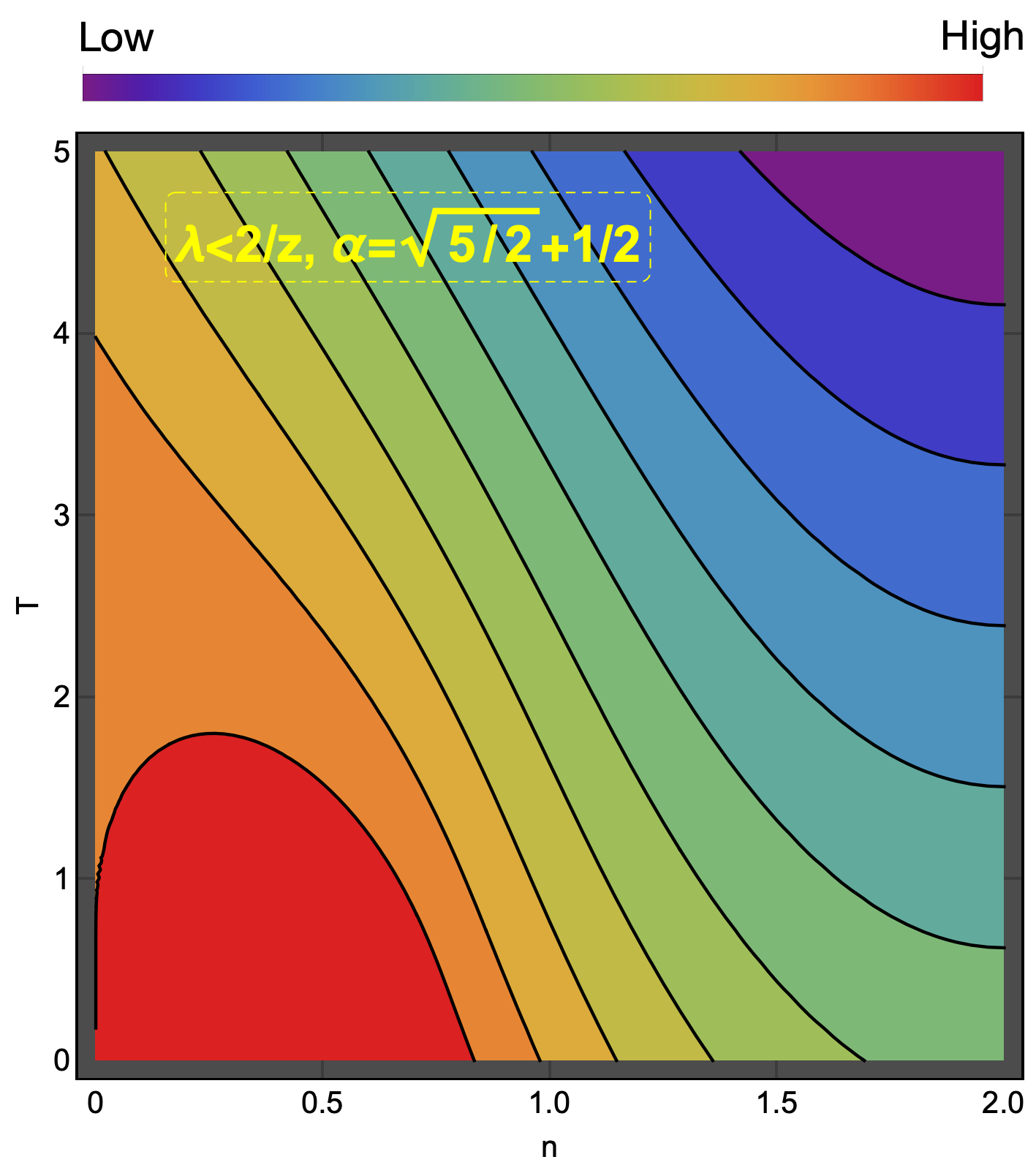}
\label{fig:FEMFLRInt-8}}
\subfigure[]{
\includegraphics[width=2.1in]{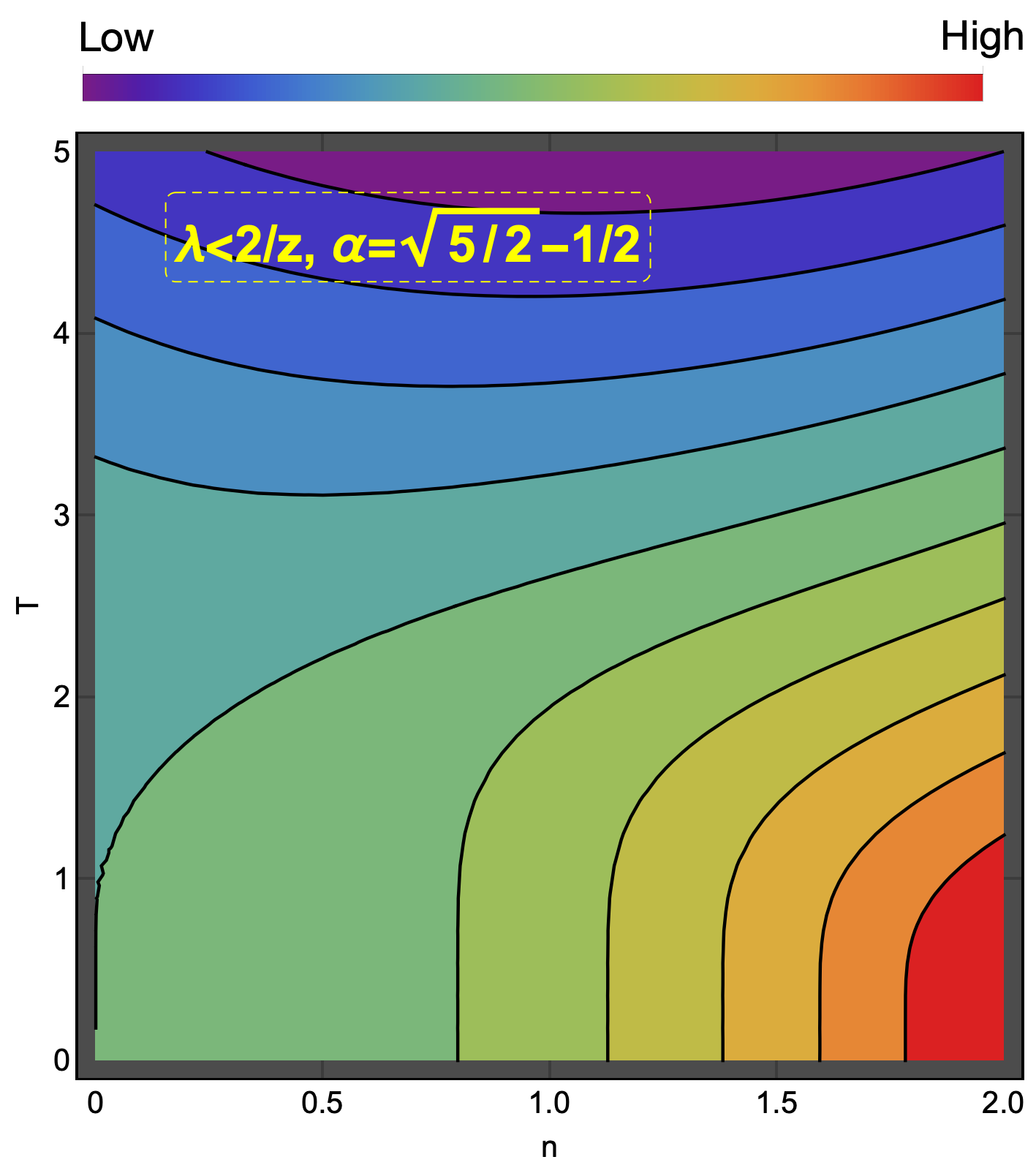}
\label{fig:FEMFLRInt-9}}
\caption{Mean field free energy per molecule contour plot as a function of temperature and spin, at low temperatures. We have fixed the values $z=6, \; \epsilon=10$.  \textbf{(a-e)} Mean field free energy considering long range interactions (Eq. \eqref{eq:FreeEnergyMF} ). We consider tuples formed by $ \lambda=\lbrace 0.25, 0.5 \rbrace$ with $ \alpha= \lbrace \sqrt{5/2}-1/2, \sqrt{5/2}, \sqrt{5/2}+1/2 \rbrace$  (see captions). \textbf{(g-i)} Mean field free energy neglecting long range interactions (Eq. \eqref{eq:FreeEnergyMF-SR}). We fixed $ \lambda=0.25$ and consider $ \alpha= \lbrace \sqrt{5/2}-1/2, \sqrt{5/2}, \sqrt{5/2}+1/2 \rbrace$  (see captions). The cases where $\lambda>2/z$ are thermodynamically prohibited since at very high temperatures the equilibrium configuration corresponds to an entropi-minimized configuration (see text for discussion), yet we consider those cases here for completeness.} \label{fig:FEMF}
\end{figure}

\end{widetext}

\begin{figure}[hbtp]
\centering
\subfigure[]{
\includegraphics[width=3.2in]{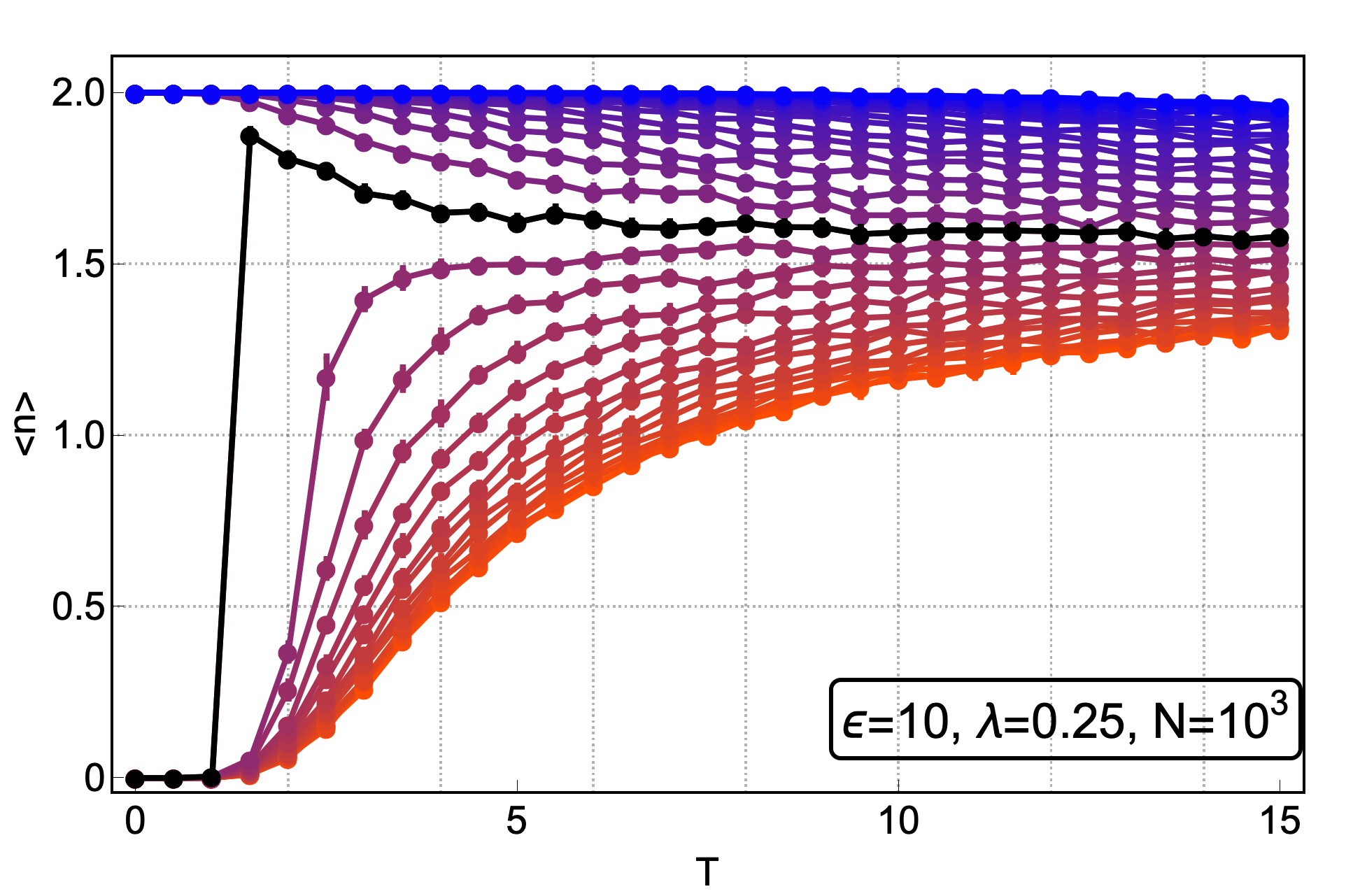}
\label{fig:Sim-Sym-a}}
\subfigure[]{
\includegraphics[width=3.2in]{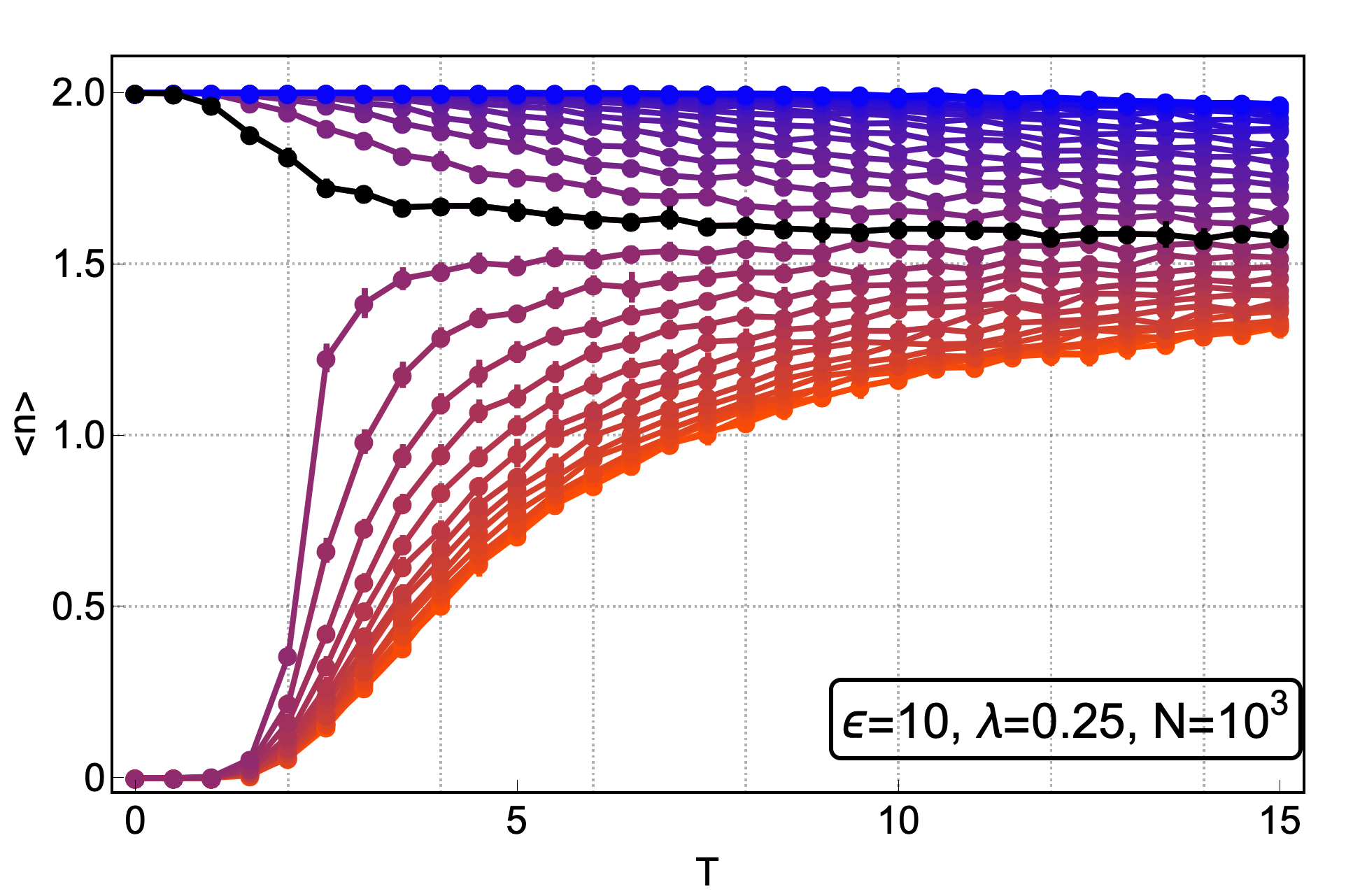}
\label{fig:Sim-Sym-b}}\\
\subfigure[]{
\includegraphics[width=3.2in]{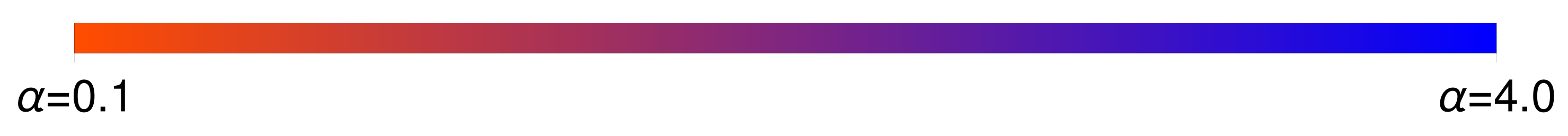}
\label{fig:Sim-bar}}
\caption{Mean spin \textit{vs} T obtained from the simulations. We have fixed the values $z=6, \; \epsilon=10$ and $\lambda=0.25$.  The black data points correspond to $\alpha=1.5$. Notice that in panel \textit{(a)} the magnetization is zero at low temperatures, while in panel \textit{(b)} the magnetization is $2$, in agreement with the mean field approximation (Eq. \eqref{eq:FreeEnergyMF}).} \label{fig:Sim-Sym}
\end{figure}

Now, in Fig. \ref{fig:FEMF} we show contour plots of the mean field Eqs. \eqref{eq:FreeEnergyMF} and \eqref{eq:FreeEnergyMF-SR} for different values of $\alpha$ and $\lambda$ in a $T$-$n$ diagram. As was discussed in the previous section,  mean field approach predicts that, in the thermodynamic limit, when long range interactions are neglected for $\alpha \approx \sqrt{5/2}$ the system is likely to be in a LS or HS state as shown in Fig. \ref{fig:FEMFLRInt-7}. In Fig. \ref{fig:Sim-Sym} we have plotted the mean spin obtained from the simulations, which we denote as, $\langle n \rangle$ \textit{vs} temperature for the aforementioned case obtained from the simulations. Notice that the black data points in Fig. \ref{fig:Sim-Sym-a}, corresponding to $\alpha=1.5$, show mean spin equal to $0$ at low temperatures. Conversely, the black data points in Fig. \ref{fig:Sim-Sym-b} show mean spin equal to $2$ at low temperatures and also correspond to $\alpha-1.5$, i.e., for the same parameter values the system is equally likely in having mean spin equal to $2$ as well as equal to $0$. This is in agreement with the mean field Eq. \eqref{eq:FreeEnergyMF} which has been plotted in Fig. \ref{fig:FEMFLRInt-7}. We have summarized all this in Table \ref{tab:table1}.

\begin{widetext}

\begin{figure}[hbtp]
\centering
\subfigure[]{
\includegraphics[width=2.2in]{Figures/SCOL025-N10.png}
\label{fig:Sim-1a}}
\subfigure[]{
\includegraphics[width=2.2in]{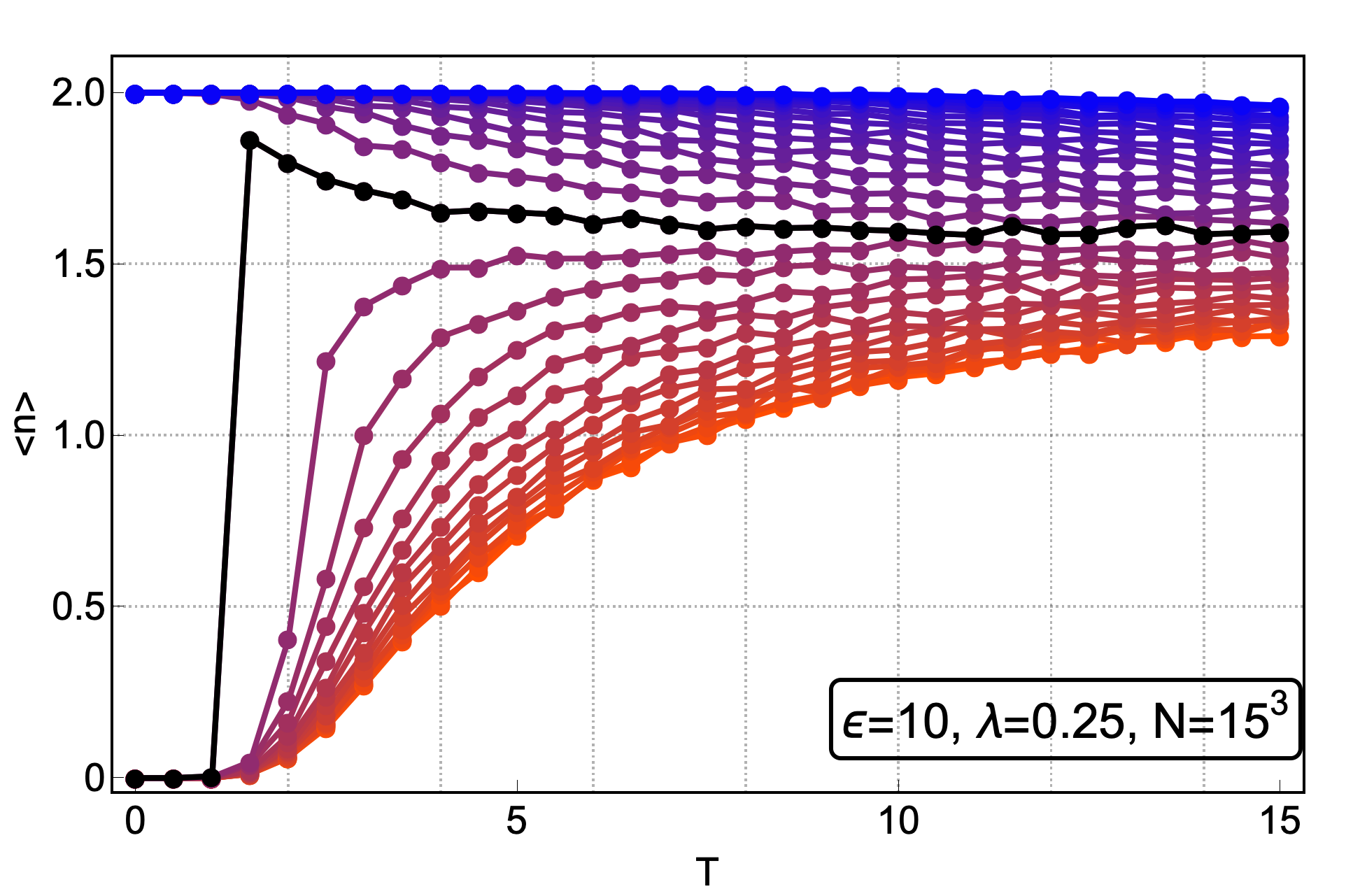}
\label{fig:Sim-1b}}
\subfigure[]{
\includegraphics[width=2.2in]{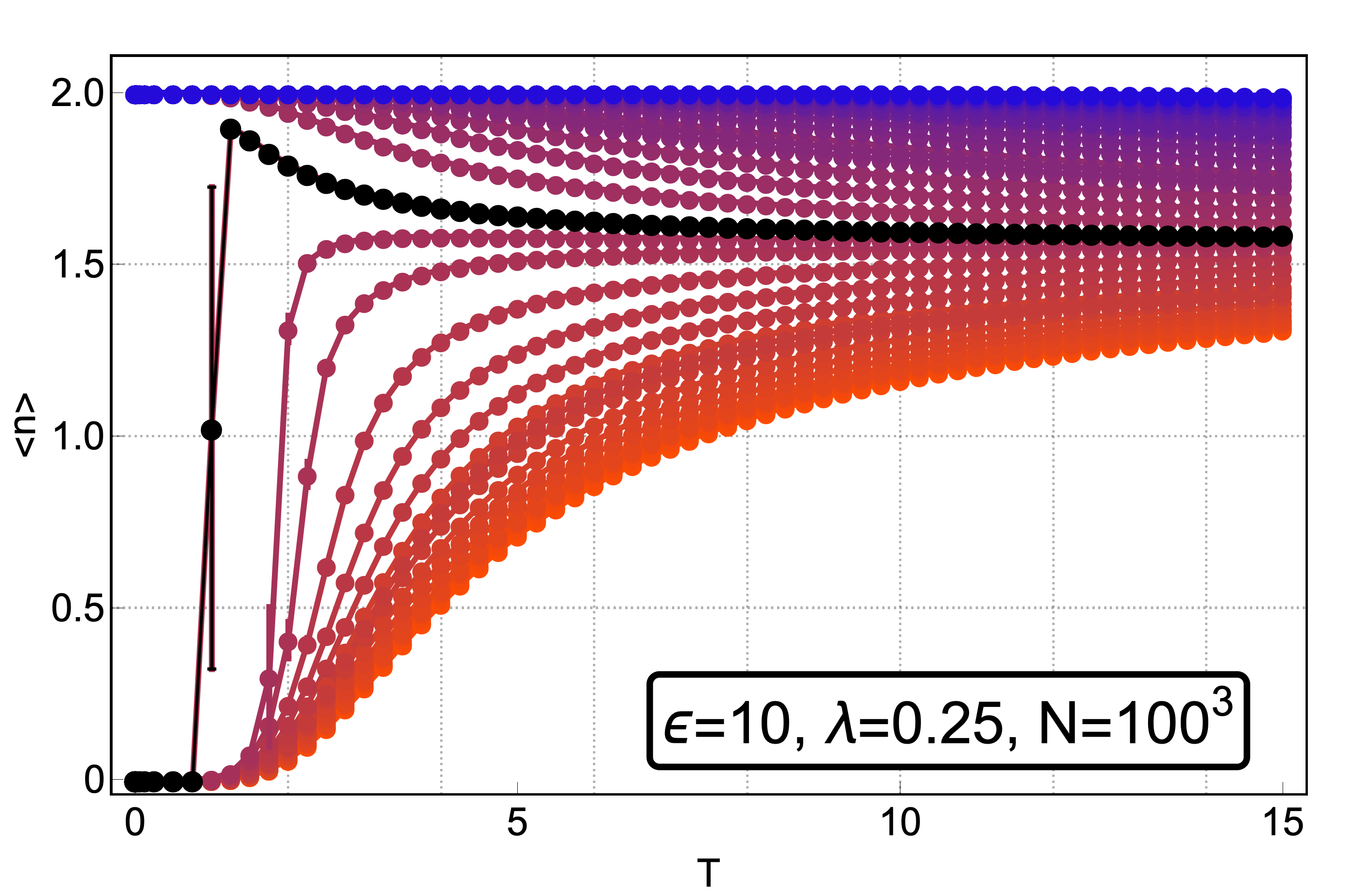}
\label{fig:Sim-1c}}
\subfigure[]{
\includegraphics[width=2.2in]{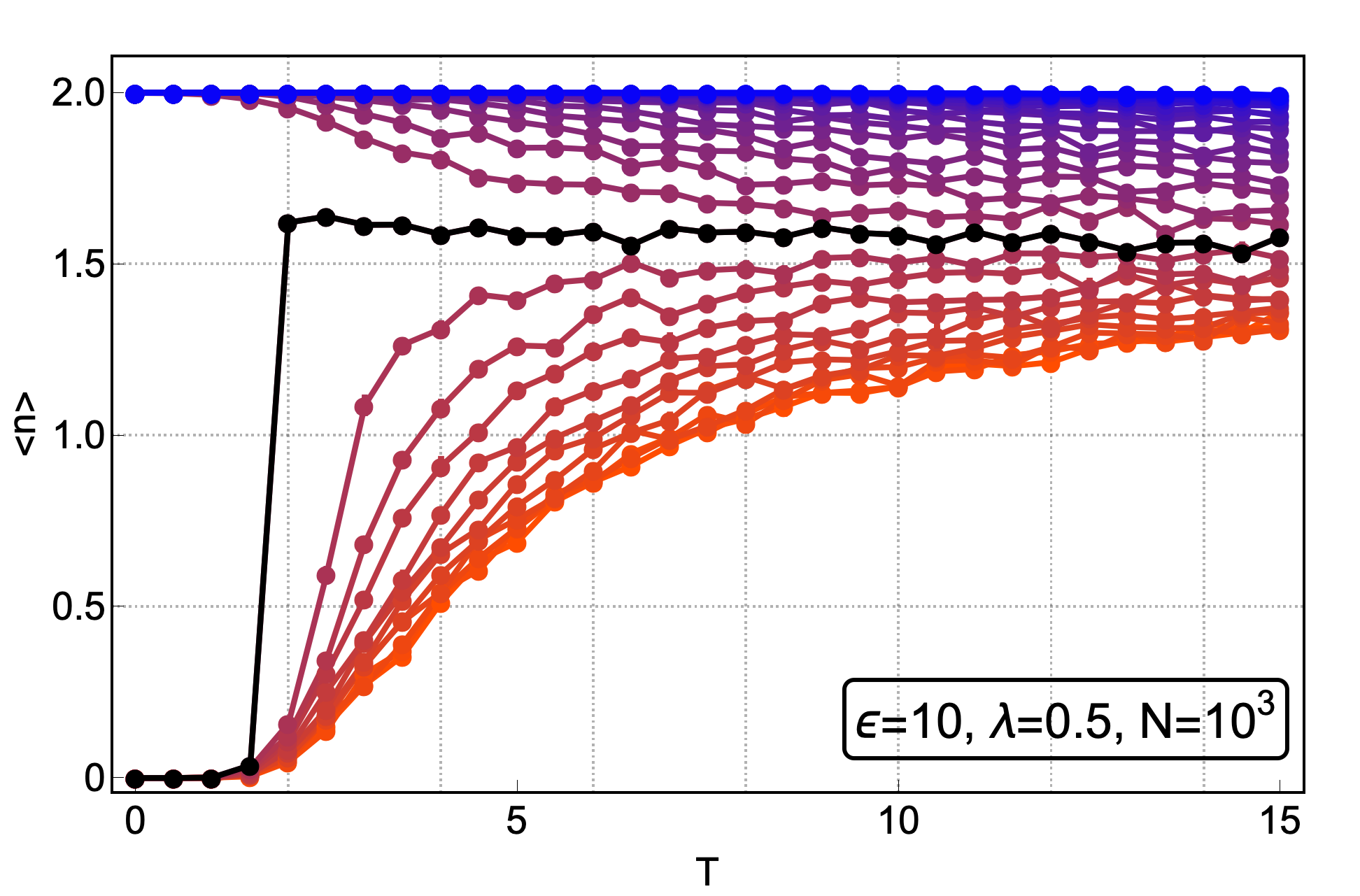}
\label{fig:Sim-2a}}
\subfigure[]{
\includegraphics[width=2.2in]{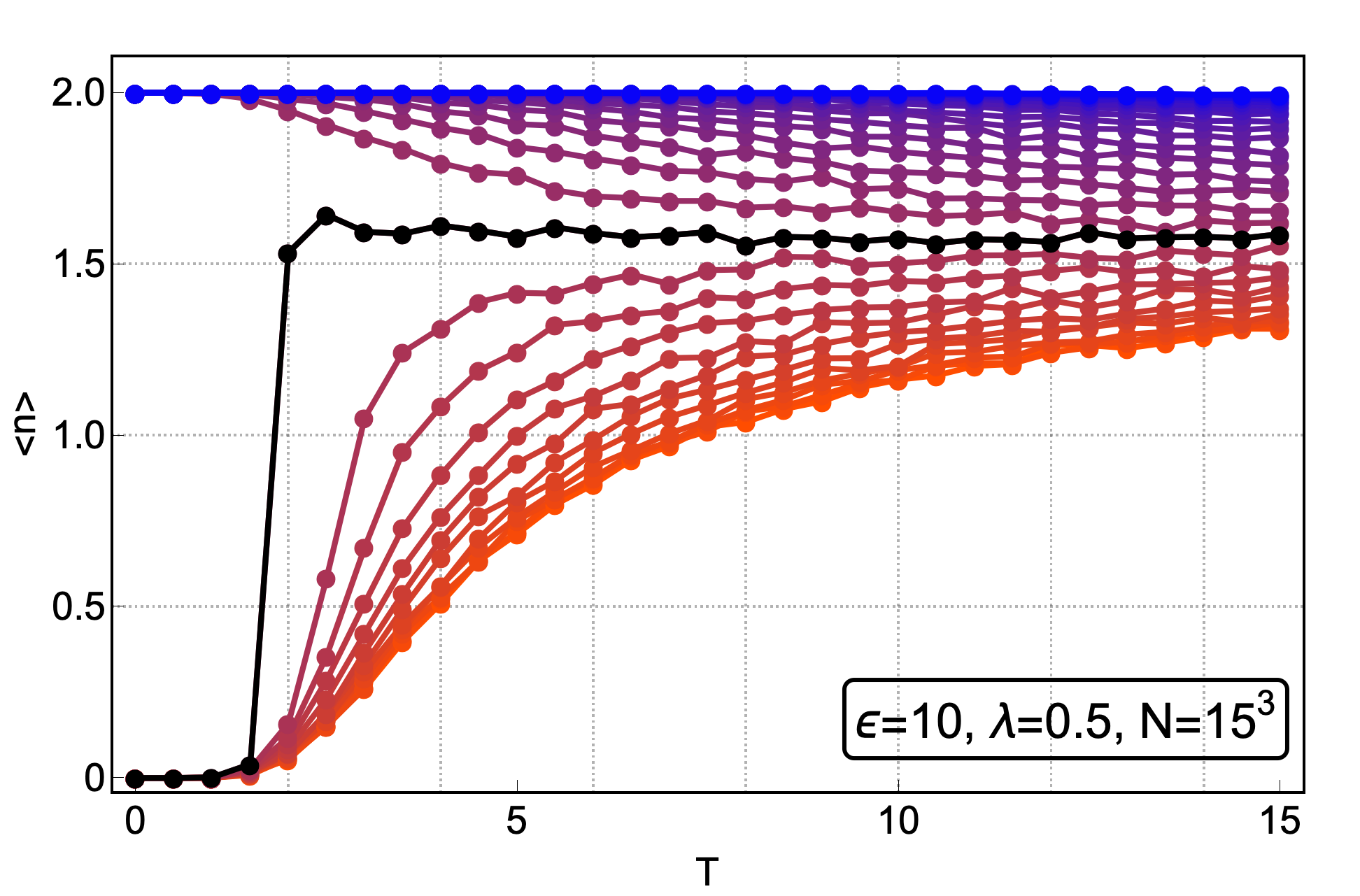}
\label{fig:Sim-2b}}
\subfigure[]{
\includegraphics[width=2.2in]{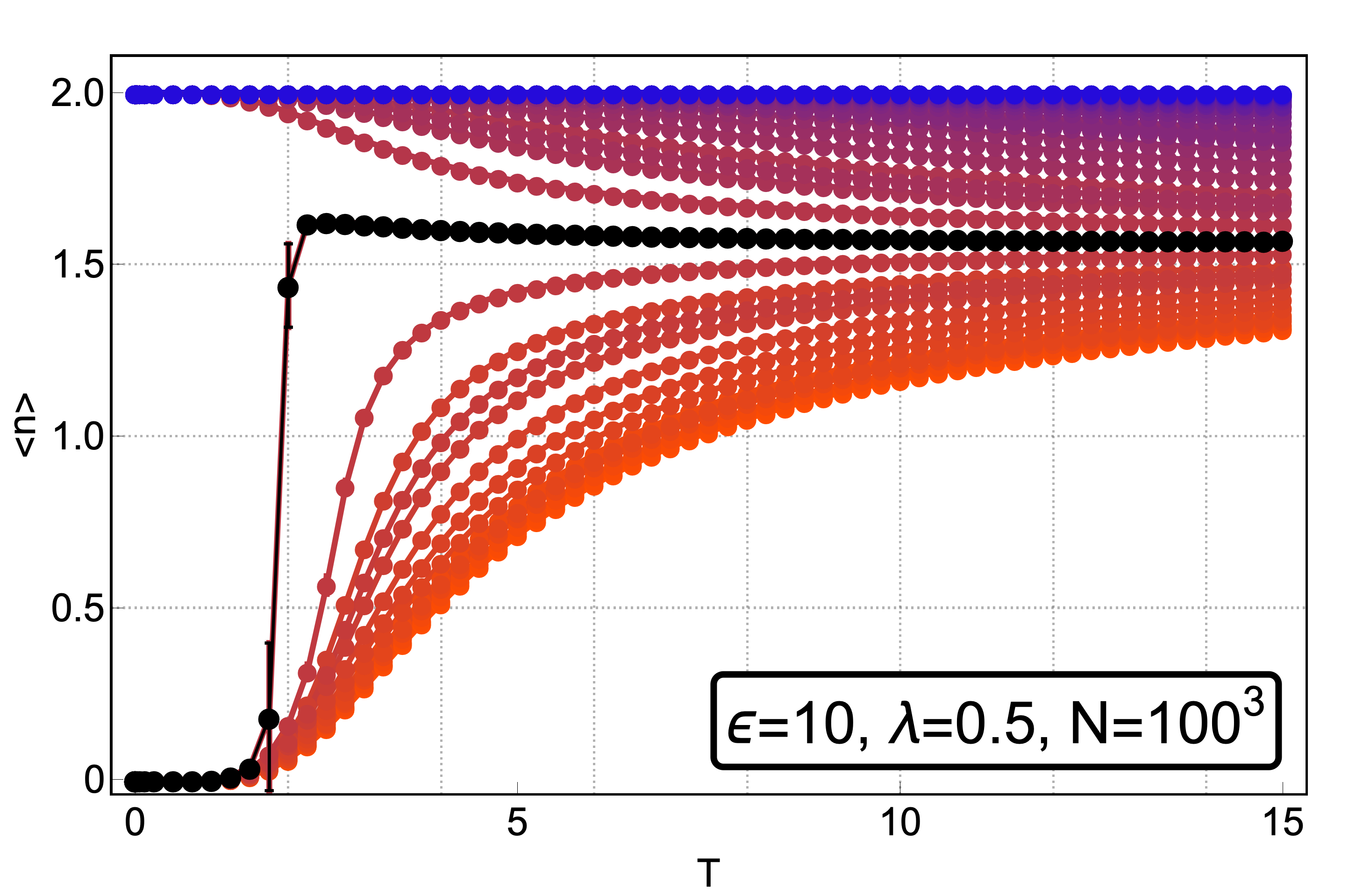}
\label{fig:Sim-2c}}
\subfigure[]{
\includegraphics[width=2.2in]{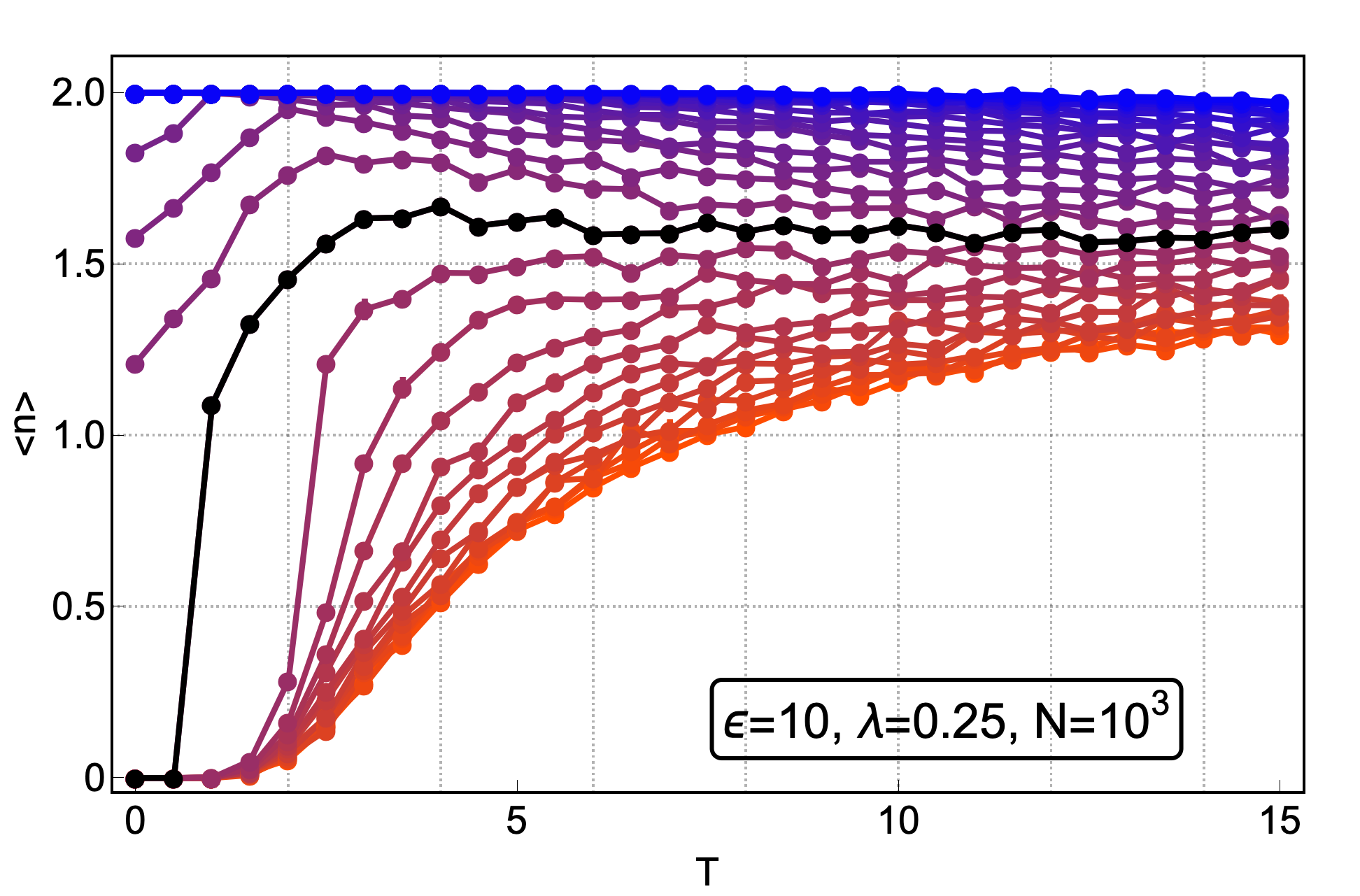}
\label{fig:Sim-3a}}
\subfigure[]{
\includegraphics[width=2.2in]{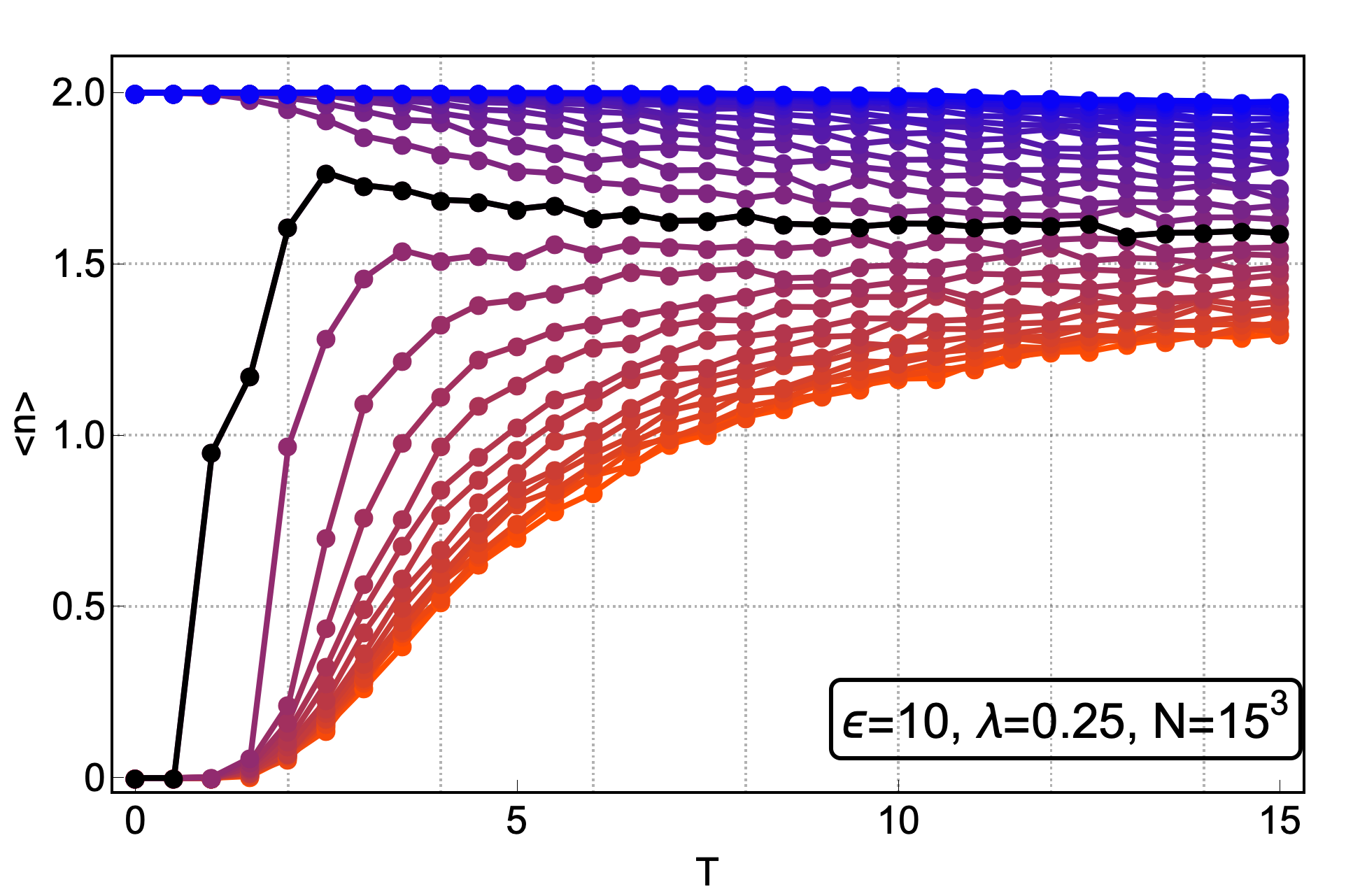}
\label{fig:Sim-3b}}
\subfigure[]{
\includegraphics[width=2.2in]{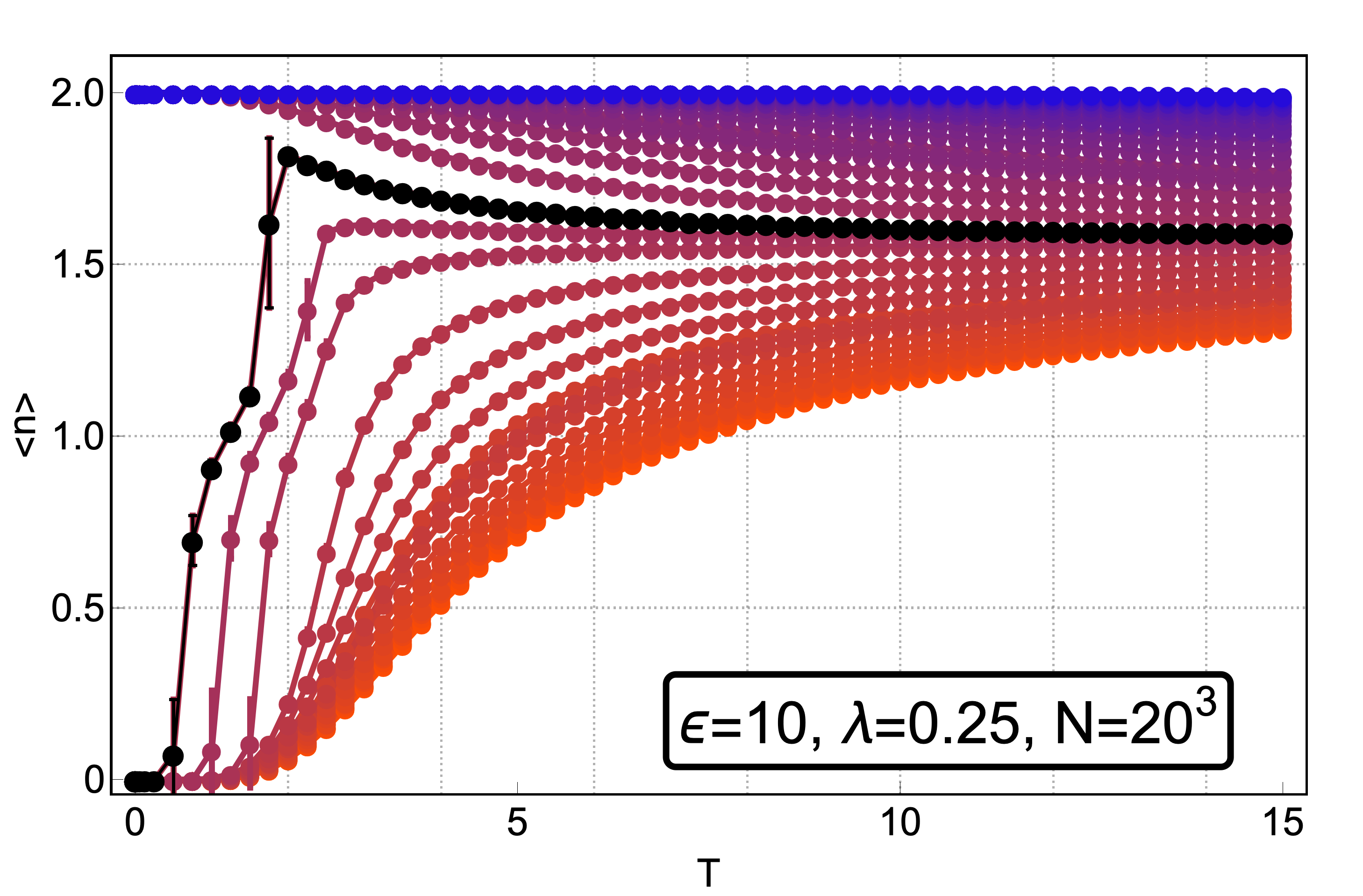}
\label{fig:Sim-3c}}
\subfigure[]{
\includegraphics[width=2.2in]{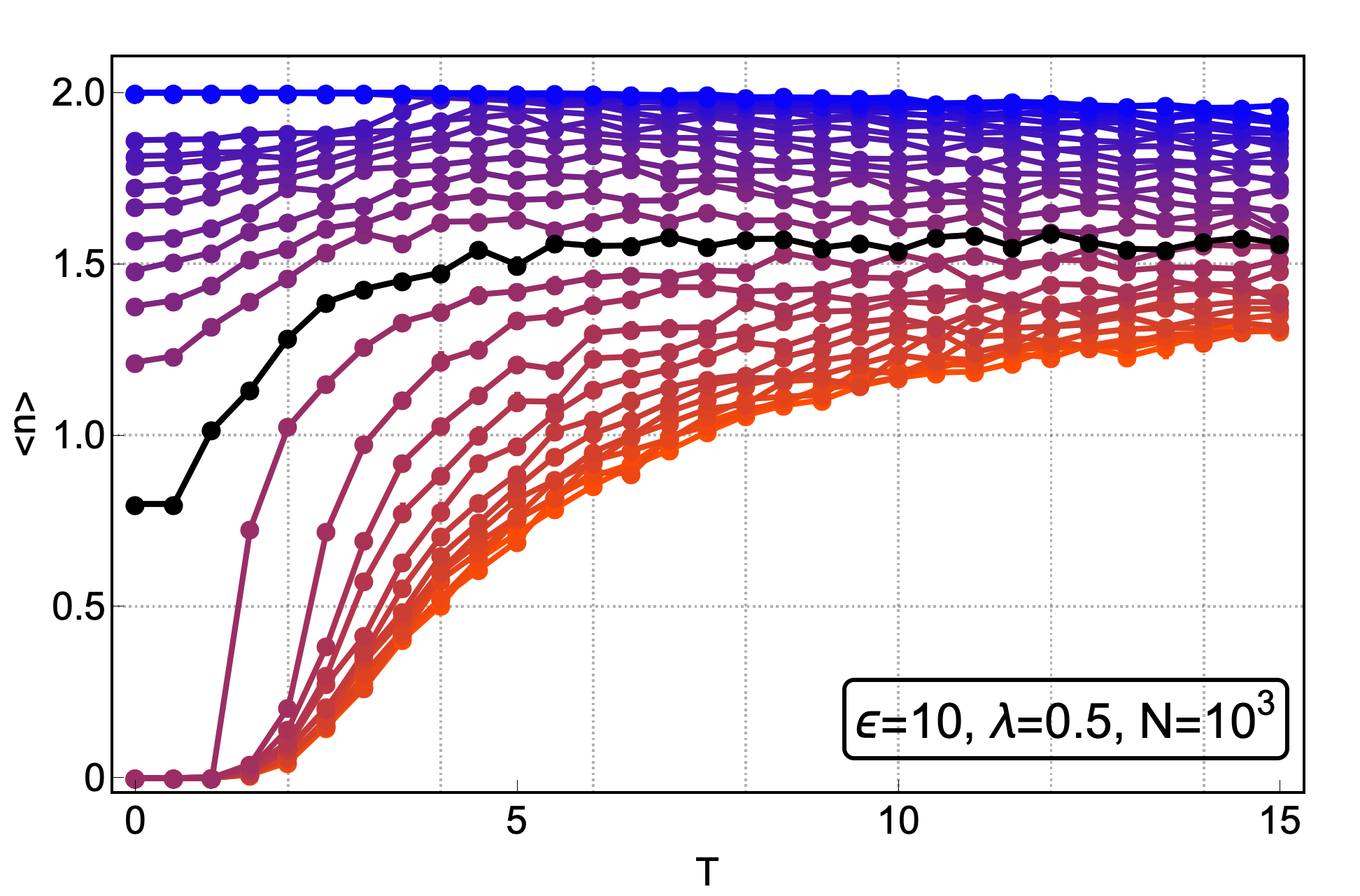}
\label{fig:Sim-4a}}
\subfigure[]{
\includegraphics[width=2.2in]{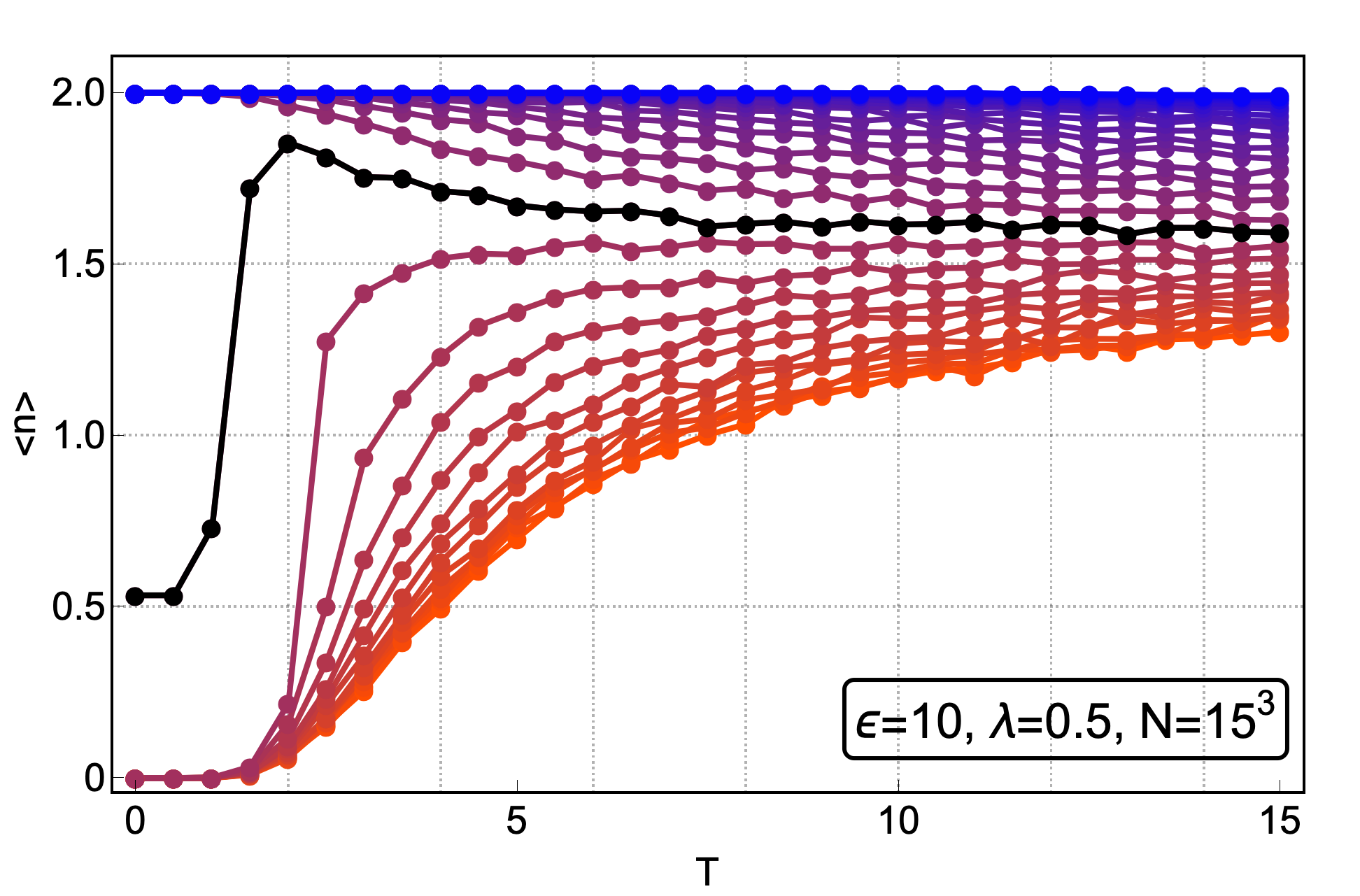}
\label{fig:Sim-4b}}
\subfigure[]{
\includegraphics[width=2.2in]{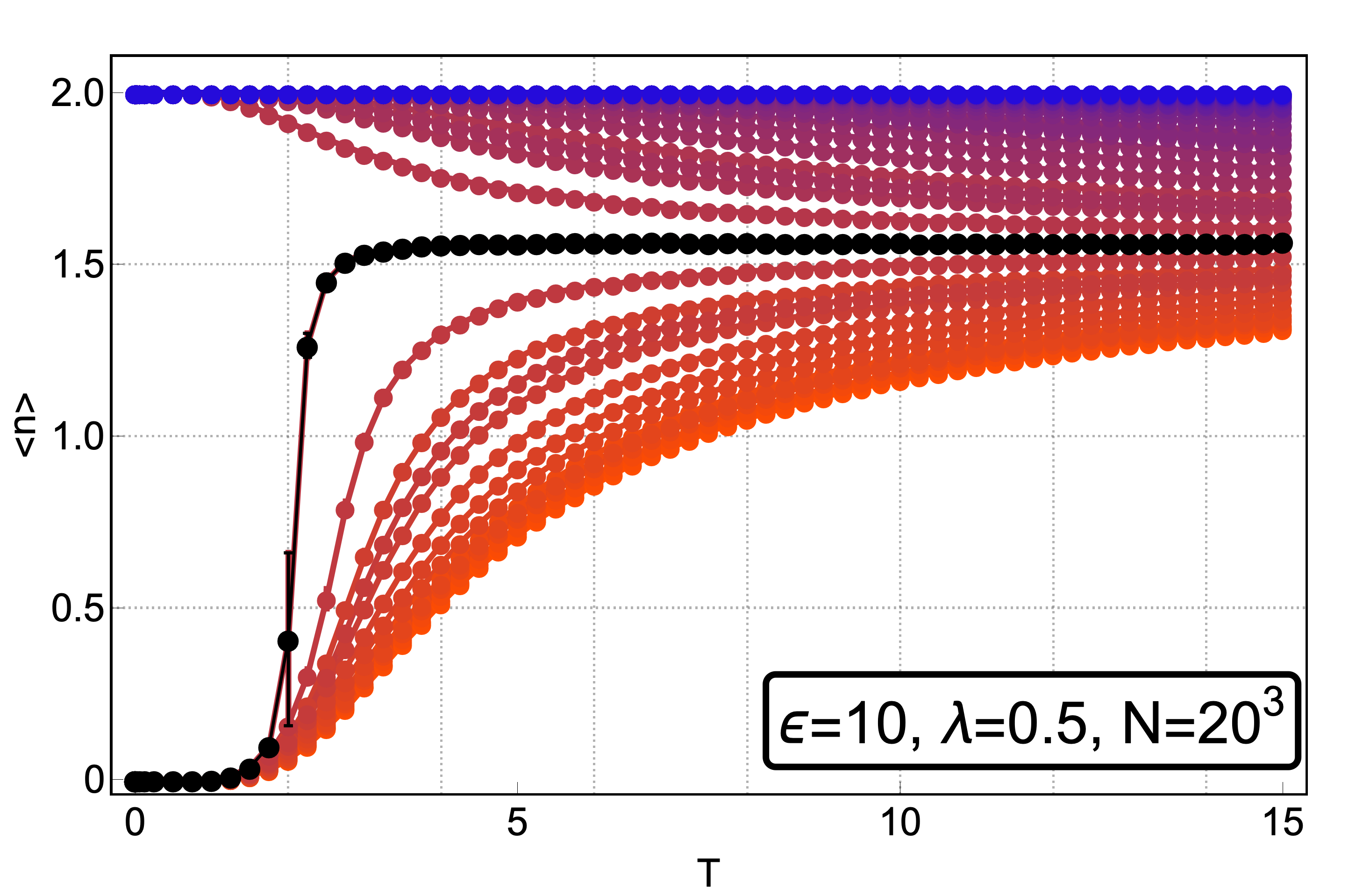}
\label{fig:Sim-4c}}\\
\subfigure[]{
\includegraphics[width=5.9in]{Figures/barlegend.png}
\label{fig:Sim-9}}
\caption{Mean spin \textit{vs} T obtained from the simulations. We have fixed the values $z=6, \; \epsilon=10$.  \textbf{(a-f)}  Long range interactions were considered for different values of $\lambda$ shown in each plot legend. Each curve corresponds to a given value of $\alpha$ which spans from $0.1$ to $4.0$. The black data points correspond to a certain $\alpha$-value in the range $\sim 1.0-1.5$ (see main text for discussion).} \label{fig:Simulations}
\end{figure}

\end{widetext}

In Fig. \ref{fig:Simulations} we show the mean spin $\langle n \rangle$ \textit{vs} temperature obtained from our simulations for different system sizes and parameter values which are specified in the legend of the figure. Fig. \ref{fig:Simulations} \textbf{(a)} to \textbf{(c)} correspond to $\epsilon=10, \; z=6$ and $\lambda=0.25$ neglecting long range interactions while considering different system sizes specified in the legends of each plot. Each curve corresponds to a fixed value of alpha that spans from $0.1$ to $4.0$ (Fig. \ref{fig:Sim-9} shows the color code we are using), as was described in detail at the beginning of this section. The first thing to notice is that for the system size we are considering, the results are robust. Secondly, notice that although initially all spins have values equal to $2$, after the system is equilibrated the mean spin value ranges from $\sim 1.25$ for low $\alpha$-value, to $2.0$ for high $\alpha$-value. Finally, the black data points correspond to $\alpha \approx 1.5$, that is the value predicted by mean field theory where the system can go to spin $0$ or spin $1$. Fig. \ref{fig:Simulations} \textbf{(d)} to \textbf{(f)} correspond to $\epsilon=10, \; z=6$ and $\lambda=0.5$ neglecting long range interactions while considering different system sizes specified in the legends of each plot. The black data points correspond to $\alpha= 1.2$. The mean field approach predicts that for $1.2<\alpha<1.3$ the system may go to spin $2$ or $0$ at low temperatures, which is consistent with these simulation results. Altogether, for the parameter values considered, the comparison between the mean field approach and the simulation results work seemingly well. Besides, the critical temperature predicted by the mean field approach is close to the values obtained in the Monte Carlo simulation, though it is advisable to not relay in mean field approaches to compute such quantities, in general.

Now, when we consider long range interactions in the Monte Carlo Simulations, the results are not as obvious as the predicted from mean field approach. Fig. \ref{fig:Simulations} \textbf{(g)} to \textbf{(i)} correspond to $\epsilon=10, \; z=6$ and $\lambda=0.25$ considering long range interactions as well as different system sizes specified in the legends of each plot. The first thing to notice is that the results obtained in the case where $N=10^3$ does not match those obtained in the case where $N=20^3$, thus a bigger system size is required for results to converge to the thermodynamical limit. Furthermore, the black data points correspond to $\alpha \approx 1.5$, i.e., around this value the system is likely to go to a spin value equal to $2$ or spin $0$. However, the mean field approach predicts a lower $\alpha$-value and this has to do with the fact that we are relying on the critical temperature obtained by the mean field approach to obtain the $\alpha$-value where the system can go to spin $2$ or spin $1$.

Finally, Fig. \ref{fig:Simulations} \textbf{(j)} to \textbf{(i)} correspond to $\epsilon=10, \; z=6$ and $\lambda=0.5$ considering long range interactions as well as different system sizes specified in the legends of each plot. Notice that the results are profoundly system size dependent, which is why mean field approaches are fruitless. Nonetheless, this is in agreement, in principle, with the theoretical prediction of the model for $\lambda< 2/z$ where it is obtained long range oscillating interactions (see SN 1). Moreover, system size dependence may be related to cooperativity properties, a feature characteristic of SCO phenomena \cite{murray2004cooperativity}. 

\section{Conclusions} \label{sec:Conc}
In this work we have studied the spin-crossover transition using a simple model derived from quite general assumptions. We have presented a thorough discussion on the parameter space using a mean field approach and Monte Carlo simulation. Using the mean field we were able to obtain the critical temperature. When long range interactions were neglected, the mean field critical temperature compared well with the Monte Carlo simulation results. We further showed, both, through the mean field approach and the Monte Carlo simulations that for a small region in the parameter space, at low temperature the model may be in LS or HS state. When long range interactions are considered, the mean field approach loses accuracy as expected since long range interactions facilitates cooperativity. We further showed that when this happens, the spin \textit{vs} $T$ curve is highly system size dependent. Hysteric effects have not been addressed here, but we may speculate that it might be possible in the parameter region where the system is likely to be in a LS or HS state. However, we leave that for future work.

\paragraph*{Acknowledgments}
We gratefully thank DGAPA-PAPIIT project IN102717. J.Q.T.M. acknowledges a doctoral fellowship from CONACyT.

\newpage

\appendix
%\numberwithin{equation}{section}
\renewcommand{\theequation}{S\arabic{equation}}

\section*{Supplementary Note 1}
Here we derive a mean field approximation for the long range interaction $U_{ij}$
First, notice that for $\bm{q} \rightarrow 0$, 
\begin{equation}
s(q\rightarrow 0)= \frac{z}{2} \; .
\end{equation}
Therefore, when $N \gg 1$, then the long range interactions become
\begin{equation}
U_{ij}=-\frac{\alpha^2 \lambda z}{2N} \frac{1}{1-\frac{2}{\lambda z}} \; . \label{eq:LRApprox2}
\end{equation}
The way Eq. \eqref{eq:LRApprox} is obtained seems rather naive. Thus, let us provide a different approach. Notice that in the case of a cubic lattice with lattice parameter $a$, in the thermodynamical limit we have
\begin{equation}
s(\bm{q})= \sum_{k=1}^d \cos (q_k) \approx \frac{z-a^2q^2}{2} \; .
\end{equation}
Thus, the long range interaction may be expressed as,
\begin{equation}
U_{ij} \approx \frac{\alpha^2 \lambda z^2}{8 \pi |\bm{r}_i - \bm{r}_j | \rho a^2} f \left(|\bm{r}_i - \bm{r}_j|, \kappa \right) \; , \label{eq:UijPot}
\end{equation}
where $f(r,\kappa)$ is defined as
\begin{equation}
f(r,\kappa)=\begin{cases}
e^{-\kappa | \bm{r}_i - \bm{r}_j | /a } \, \qquad \kappa^2 >0 \\
\cos\left(\kappa | \bm{r}_i - \bm{r}_j |/a \right) \, \qquad \kappa^2 <0
\end{cases} \; .
\end{equation}
and 
\begin{equation}
\kappa^2=\frac{2}{\lambda}-z
\end{equation}
We may further approximate this. First, for consistency, let us define $V(r_{ij})$ as
\begin{equation}
V(r_{ij})=
\begin{cases}
U_{ij} ,  \qquad r_{ij} \neq 0 \; ; \\
0 , \qquad r_{ij} =0
\end{cases} \; .
\end{equation}
Now, notice that
\begin{equation}
\sum_{i,j}V(r_{ij})= \iint d^3 \bm{r} d^3 \bm{r}' \rho(\bm{r}) \rho(\bm{r}') V(| \bm{r}- \bm{r}' |) \; ,
\end{equation}
where $\rho(\bm{r})$ is the density, i.e.,
\begin{equation}
\rho(\bm{r})= \frac{1}{V} \sum_{i=1}^N \delta \left(\bm{r} - \bm{r}_i \right) \; .
\end{equation}
Now, let us replace $\rho(\bm{r}) \rightarrow \rho \equiv N/V$ and make a change in variables to the center of mass and the relative distance taken pairwise, namely,
\begin{equation}
\begin{cases}
\bm{r}= \frac{\bm{r}_0}{2}+ \bm{R} \\
\bm{r}'= -\frac{\bm{r}_0}{2}+ \bm{R} \\
\end{cases} \; .
\end{equation}
which has Jacobian equal to unity. Henceforth, it is straightforward to show that
\begin{equation}
\sum_{i,j}V(r_{ij}) \approx \rho V 4\pi \int_0^{\infty} dr_0 r_0^2 V(r_0) \; .
\end{equation}
Applying these ideas to Eq. \eqref{eq:UijPot} yields
\begin{equation}
\sum_{i,j}U_{ij}\approx N \frac{\alpha^2 \lambda z^2}{2 a^2} \int_0^{\infty} dr_0 r_0 f(r_0,\kappa) \; .
\end{equation}
In the case $\kappa^2 >0$ the result is straightforward and yields Eq. \eqref{eq:LRApprox}. However, for $\kappa^2<0$ one should consider a maximum length. Naturally, this maximum length should be of the order of $\sim N^{1/3}$, yet, this implies that $\sum U_{ij} \sim N^{4/3}e^{N^{1/3}}$ which is fine for small system size. But the whole approximation is based on the assumption of the thermodynamical limit ($N \gg 1$). Thus, mean field does not apply in this regime.

Now, by taking into account the long range approximation, notice that
\begin{eqnarray}
\alpha^2 \left(1+\frac{\lambda z}{2} \right)N + \sum_{i,j}U_{ij}&=& \frac{\alpha^2 N}{1-\frac{\lambda z}{2}} \; , \\
\epsilon -2 n \left( \alpha^2 \left(1+\frac{\lambda z}{2} \right) + \sum_{j}U_{ij} \right) &=& \epsilon - \frac{2n \alpha^2}{1-\frac{\lambda z}{2}}
\end{eqnarray}

\section*{Supplementary Note 2}
In this section we provide a decription on how to calculate the long range interacion $U_{ij}$ in square lattices following a weighted sum is reciprocal space, using the method proposed in Ref. \cite{chadi1973special}. We need first to calculate,

\begin{equation}
s(q)=\frac{1}{2}\sum_{i(j)} \exp[i\boldsymbol{q}\cdot (\boldsymbol{R}_i-\boldsymbol{R}_j)]
\end{equation}
it follows,
\begin{equation}
s(q)=\frac{1}{2} \left[ 2 \cos(q_x a)+2 \cos(q_y a) \right]
\end{equation}
Observe that in $s(q)$, we need only to consider the lower triangle of the first quadrant where $\boldsymbol{q}$ has components $q_x>0$ and $q_y\leq q_x$, since $\cos (x)$ is an even function.

Now we need the sum, 

\begin{equation}
U_{ij}=C \sum_{\boldsymbol{q}} \frac{s^{2}(q)}{1-\lambda s(q)} \exp[i\boldsymbol{q}\cdot (\boldsymbol{R}_i-\boldsymbol{R}_j)]
\end{equation}

 Again, due to symmetry reasons, we concentrate our attention into a lower triangle of the first quadrant where $\boldsymbol{q}$ has components $q_x>0$ and $q_y\leq q_x$. 
Suppose for example that our mesh has equally spaced 16 q points. Only three of them are nonequivalent,  $\boldsymbol{q}_1=(1/8,1/8)$ appears 4 times, $\boldsymbol{q}_2=(3/8,3/8)$ appears 4 times and $\boldsymbol{q}_3=(3/8,1/8)$ 8 times. Thus
each point in a sum has a weight, 
\begin{equation}
W_1=1/4, W_2=1/4,W_3=1/2
\end{equation}

We now use our mesh in $\bm{q}$-space,
\begin{equation}
U_{ij}= \sum_{l=1}^{3} W_l F(l) \cos[\boldsymbol{q}_l\cdot (\boldsymbol{R}_i-\boldsymbol{R}_j)]
\end{equation}
where,
\begin{equation}
F(l)= \frac{s^{2}(q_l)}{1-\lambda s(q_l)}
\end{equation}
or,

\begin{equation}
U_{ij}= \sum_{l=1}^{3} W_l F(l) cos[q_l R_{ij} \cos(\theta_l-\theta_{ij})]
\end{equation}
where $\theta_l=\tan^{-1}(q_l^{x}/q_l^{y})$ and $q_l=\sqrt{(q_l^{x})^{2}+(q_l^{y})^{2}}$

The mesh can be further refined using this technique. 

Let us calculate $U_{ij}$ in a cubic lattice. First,
\begin{equation}
s(q)=\frac{1}{2} \left[ 2\cos(q_x a)+2\cos(q_y a) +2\cos(q_za)\right]
\end{equation}

In a 3D cubic lattice, we use a mesh grid of 64 $q$ points. This can be reduced to compute only these $q$-points \cite{chadi1973special},\\
$\boldsymbol{q}_1=(1/8,1/8,1/8)$ with $W_1=1/8$\\
  $\boldsymbol{q}_2=(3/8,3/8,1/8)$  and $W_2=3/8$\\ 
  $\boldsymbol{q}_3=(3/8,1/8,1/8)$  and $W_3=3/8$\\
$\boldsymbol{q}_4=(3/8,3/8,3/8)$ , and $W_4=1/8$\\

\begin{equation}
U_{ij}= \sum_{l=1}^{4} W_l F(l) \cos[q_l R_{ij} \cos(\theta_l-\theta_{ij})]
\end{equation}

If required, consider points
\begin{equation}
\boldsymbol{q}_l=\frac{1}{8}(1/2^{n},1/2^{n},1/2^{n})
\end{equation}
with $n$ an integer and then iterate. Then,
\begin{equation}
W_l=\frac{n_l}{\sum_j n_j}
\end{equation}
where $n_l$ is the number of points obtained from $\boldsymbol{q}_l$ under all
symmetry operations of the lattice point group.  

\section*{Supplementary Note 3}
In this section we show how to obtain the effective Hamiltonian (Eq. \eqref{eq:HamElec1}). To this end, we consider a $d$-dimensional periodic array of $N$ mono-nuclear metal complexes which we tag with latin letters. Each site $i$ contains an ion $\text{Fe}^{+2}$ in an octahedral site surrounded by non-magnetic ligands. We denote the number of electrons occupying states eg at ion $i$th with $n_i$. These electrons are coupled to a local “breathing” vibration mode described by the creation and annihilation operators over site $i$ denoted as $\hat{a}_i^{\dag}$ and $\hat{a}_i$, respectively. Breathing modes at neighboring sites are also coupled through their interaction with acoustic phonons. Thus, the effective Hamiltonian for this electron-local vibrations system may be expressed as:
\begin{equation}
\hat{H}=\sum_{i=1}^N \hat{H}_i + \sum_{\langle i,j \rangle} \hat{V}_{ij} \; , \label{eq:HamHam}
\end{equation}
with
\begin{equation}
\begin{cases}
\hat{H}_i=\epsilon n_i+\left(\hat{a}_i^{\dag} \hat{a}_i + \frac{1}{2} \right) - \alpha n_i \left(\hat{a}_i^{\dag}+\hat{a}_i \right) \; , \\
\hat{V}_{ij}= -\frac{\lambda}{4} \left(\hat{a}_i^{\dag} + \hat{a}_i \right) \left(\hat{a}_j^{\dag} + \hat{a}_j \right) \; .
\end{cases} 
\end{equation}
In these expressions $\epsilon$ is the excitation energy per eg electron, while $\alpha$ and $\lambda$ are coupling parameters. 
Now, let us consider a transformation on the local phonon operators, such that,
\begin{equation}
\begin{cases}
 \hat{a}_i = \hat{b}_i +\alpha n_i \; , \\
  \hat{a}_i^{\dag}= \hat{b}_i^{\dag}+ \alpha n_i \; . 
 \end{cases}
\end{equation}
Therefore, Eq. \eqref{eq:HamHam} becomes,
\begin{eqnarray}
&&\hat{H}= \sum_{i=1}^N \left(\epsilon n_i - \alpha^2 n_i^2 + \hat{b}_i^{\dag}  \hat{b}_i + \frac{1}{2}  \right) - \alpha^2 \lambda \sum_{\langle i , j \rangle} n_i n_j \nonumber \\
&& - \alpha \lambda \sum_{\langle i,j \rangle} \left( \hat{b}_i^{\dag} +  \hat{b}_i \right)n_j -\frac{\lambda}{4} \sum_{\langle i,j \rangle} \left( \hat{b}_i^{\dag} +  \hat{b}_i \right) \left( \hat{b}_j^{\dag} +  \hat{b}_j \right) .
\end{eqnarray}
Now, let us express the operators $\hat{b}_i^{\dag}$ and $\hat{b}_i$ in Fourier space, i.e.,
\begin{equation}
\begin{cases}
\hat{b}_i^{\dag} = \frac{1}{\sqrt{N}} \sum_{\bm{q}} \hat{b}_{\bm{q}}^{\dag} e^{- \imath \bm{q} \cdot \bm{r}_i} \; , \\
\hat{b}_i = \frac{1}{\sqrt{N}} \sum_{\bm{q}} \hat{b}_{\bm{q}} e^{- \imath \bm{q} \cdot \bm{r}_i}
\end{cases} \; .
\end{equation}
Notice that
\begin{equation}
\sum_{i} \hat{b}_i^{\dag} \hat{b}_i = \sum_{\bm{q}} \hat{b}_{\bm{q}}^{\dag} \hat{b}_{\bm{q}} \; ,
\end{equation}
where we have used the identity,
\begin{equation}
\delta_{\bm{q} \bm{q}'} = \frac{1}{N} \sum_{i} e^{-\imath \left(\bm{q} - \bm{q}' \right) \cdot \bm{r}_i} \; .
\end{equation}
Here $\delta_{\bm{q} \bm{q}'}$ is the Kronecker delta defined as
\begin{equation}
\delta_{\bm{q} \bm{q}'}=
\begin{cases}
1, \quad \bm{q}=\bm{q}' \\
0, \quad \bm{q} \neq \bm{q}'
\end{cases} \; .
\end{equation}
Now, let us define $s(\bm{q})$ as
\begin{equation}
s(\bm{q})=\frac{1}{2} \sum_{j(i)} e^{\imath \bm{q} \cdot \left(\bm{r}_j - \bm{r}_i \right)} \; .
\end{equation}
where we denote $j(i)$ as the index $j$ running over the first-neighbor lattice sites of site $i$. Now, for a lattice with inversion symmetry, i.e., $s(\bm{q})=s(-\bm{q})$, the phonon part of the Hamiltonian, $\hat{H}_{ph}$, becomes
\begin{eqnarray}
\hat{H}_{ph} &=& \sum_{i=1}^N \left(\hat{b}_i^{\dag}  \hat{b}_i + \frac{1}{2} \right) \nonumber -\frac{\lambda}{4} \sum_{\langle i,j \rangle} \left( \hat{b}_i^{\dag} +  \hat{b}_i \right) \left( \hat{b}_j^{\dag} +  \hat{b}_j \right) \\
&=& \sum_{\bm{q}} \left(\hat{b}_{\bm{q}}^{\dag} \hat{b}_{\bm{q}} + \frac{1}{2} \right. \nonumber \\
&&\left. - \frac{\lambda}{4} s(\bm{q}) \left( \hat{b}_{\bm{q}}^{\dag} \hat{b}_{\bm{-q}}^{\dag} + \hat{b}_{\bm{q}}^{\dag} \hat{b}_{\bm{q}} + \hat{b}_{\bm{q}} \hat{b}_{\bm{q}}^{\dag} + \hat{b}_{\bm{q}} \hat{b}_{\bm{-q}} \right) \right) \label{eq:HamPh}
\; .
\end{eqnarray}
In order to express Eq. \eqref{eq:HamPh} in a canonical way, we further introduce the new operators $\hat{f}_{\bm{q}}^{\dag}$ and $\hat{f}_{\bm{q}}$, such that,
\begin{equation}
\begin{cases}
\hat{b}_{\bm{q}} = \cosh \left(u (\bm{q}) \right) \hat{f}_{\bm{q}} + \sinh \left(u (\bm{q}) \right)  \hat{f}_{-\bm{q}}^{\dag} \; , \\
\hat{b}_{\bm{q}}^{\dag} = \cosh \left(u (\bm{q}) \right) \hat{f}_{\bm{q}}^{\dag} + \sinh \left(u (\bm{q}) \right)  \hat{f}_{-\bm{q}} \; 
\end{cases} \; . \label{eq:Trans2}
\end{equation}

Substituting Eq. \eqref{eq:Trans2} in Eq. \eqref{eq:HamPh} and after some algebra, one obtains
\begin{equation}
\hat{H}_{ph} = \sum_{\bm{q}} \omega (\bm{q}) \left(\hat{f}_{\bm{q}}^{\dag} \hat{f}_{\bm{q}} + \frac{1}{2} \right) \; ,
\end{equation}
with
\begin{equation}
\omega (\bm{q}) = \sqrt{1 - \lambda s(\bm{q})} \; .
\end{equation}
Similarly, the electron-phonon coupled term may be decoupled by introducing the same transformations. It is not difficult to show that,
\begin{eqnarray}
&&\hat{H}_{ep}= -\alpha \lambda \sum_{\langle i,j \rangle} \left(\hat{b}_{i}^{\dag} + \hat{b}_{i} \right)n_j \nonumber \\
&&= - \frac{\alpha \lambda}{\sqrt{N}} \sum_{\bm{q},i} s(\bm{q}) e^{u \left(\bm{q} \right)} \left(\hat{f}_{\bm{q}}^{\dag} e^{-\imath \bm{q} \cdot \bm{r}_i} + \hat{f}_{\bm{q}} e^{\imath \bm{q} \cdot \bm{r}_i} \right)n_i \; .
\end{eqnarray}
Finally, we define $ \hat{c}_{\bm{q}}^{\dag}$ and  $\hat{c}_{\bm{q}}$ as,
\begin{equation}
\begin{cases}
 \hat{c}_{\bm{q}}^{\dag} =  \hat{f}_{\bm{q}}^{\dag} - \frac{\lambda \alpha s(\bm{q}) e^{u(\bm{q})}}{\omega(\bm{q}) \sqrt{N}} \sum_{i} n_i e^{-\imath \bm{q} \cdot \bm{r}_i} \; , \\
  \hat{c}_{\bm{q}} =  \hat{f}_{\bm{q}} - \frac{\lambda \alpha s(\bm{q}) e^{u(\bm{q})}}{\omega(\bm{q}) \sqrt{N}} \sum_{i} n_i e^{\imath \bm{q} \cdot \bm{r}_i} \; ,
\end{cases}
\end{equation}
 and arrange terms to finally obtain,
 \begin{eqnarray}
 \hat{H} &=& \sum_{i=1}^N \left(\epsilon n_i - \alpha^2 n_i^2 \right) - \alpha^2 \lambda \sum_{\langle i,j \rangle} n_i n_j - \sum_{i,j} U_{ij}n_i n_j \nonumber \\
 &&+ \sum_{\bm{q}} \omega(\bm{q}) \left( \hat{c}_{\bm{q}}^{\dag}  \hat{c}_{\bm{q}} + \frac{1}{2} \right) \; .
 \end{eqnarray}

\bibliography{mybib}

\begin{thebibliography}{22}%
\makeatletter
\providecommand \@ifxundefined [1]{%
 \@ifx{#1\undefined}
}%
\providecommand \@ifnum [1]{%
 \ifnum #1\expandafter \@firstoftwo
 \else \expandafter \@secondoftwo
 \fi
}%
\providecommand \@ifx [1]{%
 \ifx #1\expandafter \@firstoftwo
 \else \expandafter \@secondoftwo
 \fi
}%
\providecommand \natexlab [1]{#1}%
\providecommand \enquote  [1]{``#1''}%
\providecommand \bibnamefont  [1]{#1}%
\providecommand \bibfnamefont [1]{#1}%
\providecommand \citenamefont [1]{#1}%
\providecommand \href@noop [0]{\@secondoftwo}%
\providecommand \href [0]{\begingroup \@sanitize@url \@href}%
\providecommand \@href[1]{\@@startlink{#1}\@@href}%
\providecommand \@@href[1]{\endgroup#1\@@endlink}%
\providecommand \@sanitize@url [0]{\catcode `\\12\catcode `\$12\catcode
  `\&12\catcode `\#12\catcode `\^12\catcode `\_12\catcode `\%12\relax}%
\providecommand \@@startlink[1]{}%
\providecommand \@@endlink[0]{}%
\providecommand \url  [0]{\begingroup\@sanitize@url \@url }%
\providecommand \@url [1]{\endgroup\@href {#1}{\urlprefix }}%
\providecommand \urlprefix  [0]{URL }%
\providecommand \Eprint [0]{\href }%
\providecommand \doibase [0]{http://dx.doi.org/}%
\providecommand \selectlanguage [0]{\@gobble}%
\providecommand \bibinfo  [0]{\@secondoftwo}%
\providecommand \bibfield  [0]{\@secondoftwo}%
\providecommand \translation [1]{[#1]}%
\providecommand \BibitemOpen [0]{}%
\providecommand \bibitemStop [0]{}%
\providecommand \bibitemNoStop [0]{.\EOS\space}%
\providecommand \EOS [0]{\spacefactor3000\relax}%
\providecommand \BibitemShut  [1]{\csname bibitem#1\endcsname}%
\let\auto@bib@innerbib\@empty
%</preamble>
\bibitem [{\citenamefont {Halcrow}(2013)}]{halcrow2013spin}%
  \BibitemOpen
  \bibfield  {author} {\bibinfo {author} {\bibfnamefont {M.~A.}\ \bibnamefont
  {Halcrow}},\ }\href@noop {} {\emph {\bibinfo {title} {Spin-crossover
  materials: properties and applications}}}\ (\bibinfo  {publisher} {John Wiley
  \& Sons},\ \bibinfo {year} {2013})\BibitemShut {NoStop}%
\bibitem [{\citenamefont {G{\"u}tlich}\ and\ \citenamefont
  {Goodwin}(2004)}]{gutlich2004spin}%
  \BibitemOpen
  \bibfield  {author} {\bibinfo {author} {\bibfnamefont {P.}~\bibnamefont
  {G{\"u}tlich}}\ and\ \bibinfo {author} {\bibfnamefont {H.~A.}\ \bibnamefont
  {Goodwin}},\ }in\ \href@noop {} {\emph {\bibinfo {booktitle} {Spin Crossover
  in Transition Metal Compounds I}}}\ (\bibinfo  {publisher} {Springer},\
  \bibinfo {year} {2004})\ pp.\ \bibinfo {pages} {1--47}\BibitemShut {NoStop}%
\bibitem [{\citenamefont {Halcrow}(2011)}]{halcrow2011structure}%
  \BibitemOpen
  \bibfield  {author} {\bibinfo {author} {\bibfnamefont {M.~A.}\ \bibnamefont
  {Halcrow}},\ }\href@noop {} {\bibfield  {journal} {\bibinfo  {journal}
  {Chemical Society Reviews}\ }\textbf {\bibinfo {volume} {40}},\ \bibinfo
  {pages} {4119} (\bibinfo {year} {2011})}\BibitemShut {NoStop}%
\bibitem [{\citenamefont {Shriver}, \citenamefont {Atkins},\ and\ \citenamefont
  {Langford}(1994)}]{shriver1994standard}%
  \BibitemOpen
  \bibfield  {author} {\bibinfo {author} {\bibfnamefont {D.}~\bibnamefont
  {Shriver}}, \bibinfo {author} {\bibfnamefont {P.}~\bibnamefont {Atkins}}, \
  and\ \bibinfo {author} {\bibfnamefont {C.}~\bibnamefont {Langford}},\
  }\href@noop {} {\bibfield  {journal} {\bibinfo  {journal} {Inorganic
  Chemistry}\ ,\ \bibinfo {pages} {B7}} (\bibinfo {year} {1994})}\BibitemShut
  {NoStop}%
\bibitem [{\citenamefont {G{\"u}tlich}, \citenamefont {Hauser},\ and\
  \citenamefont {Spiering}(1994)}]{gutlich1994thermal}%
  \BibitemOpen
  \bibfield  {author} {\bibinfo {author} {\bibfnamefont {P.}~\bibnamefont
  {G{\"u}tlich}}, \bibinfo {author} {\bibfnamefont {A.}~\bibnamefont {Hauser}},
  \ and\ \bibinfo {author} {\bibfnamefont {H.}~\bibnamefont {Spiering}},\
  }\href@noop {} {\bibfield  {journal} {\bibinfo  {journal} {Angewandte Chemie
  International Edition in English}\ }\textbf {\bibinfo {volume} {33}},\
  \bibinfo {pages} {2024} (\bibinfo {year} {1994})}\BibitemShut {NoStop}%
\bibitem [{\citenamefont {Yang}\ \emph {et~al.}(2015)\citenamefont {Yang},
  \citenamefont {Tong}, \citenamefont {Lin}, \citenamefont {Okuchi},\ and\
  \citenamefont {Tomioka}}]{yang2015elasticity}%
  \BibitemOpen
  \bibfield  {author} {\bibinfo {author} {\bibfnamefont {J.}~\bibnamefont
  {Yang}}, \bibinfo {author} {\bibfnamefont {X.}~\bibnamefont {Tong}}, \bibinfo
  {author} {\bibfnamefont {J.-F.}\ \bibnamefont {Lin}}, \bibinfo {author}
  {\bibfnamefont {T.}~\bibnamefont {Okuchi}}, \ and\ \bibinfo {author}
  {\bibfnamefont {N.}~\bibnamefont {Tomioka}},\ }\href@noop {} {\bibfield
  {journal} {\bibinfo  {journal} {Scientific reports}\ }\textbf {\bibinfo
  {volume} {5}},\ \bibinfo {pages} {17188} (\bibinfo {year}
  {2015})}\BibitemShut {NoStop}%
\bibitem [{\citenamefont {L{\'e}tard}, \citenamefont {Guionneau},\ and\
  \citenamefont {Goux-Capes}(2004)}]{letard2004towards}%
  \BibitemOpen
  \bibfield  {author} {\bibinfo {author} {\bibfnamefont {J.-F.}\ \bibnamefont
  {L{\'e}tard}}, \bibinfo {author} {\bibfnamefont {P.}~\bibnamefont
  {Guionneau}}, \ and\ \bibinfo {author} {\bibfnamefont {L.}~\bibnamefont
  {Goux-Capes}},\ }in\ \href@noop {} {\emph {\bibinfo {booktitle} {Spin
  Crossover in Transition Metal Compounds III}}}\ (\bibinfo  {publisher}
  {Springer},\ \bibinfo {year} {2004})\ pp.\ \bibinfo {pages}
  {221--249}\BibitemShut {NoStop}%
\bibitem [{\citenamefont {Halder}\ \emph {et~al.}(2002)\citenamefont {Halder},
  \citenamefont {Kepert}, \citenamefont {Moubaraki}, \citenamefont {Murray},\
  and\ \citenamefont {Cashion}}]{halder2002guest}%
  \BibitemOpen
  \bibfield  {author} {\bibinfo {author} {\bibfnamefont {G.~J.}\ \bibnamefont
  {Halder}}, \bibinfo {author} {\bibfnamefont {C.~J.}\ \bibnamefont {Kepert}},
  \bibinfo {author} {\bibfnamefont {B.}~\bibnamefont {Moubaraki}}, \bibinfo
  {author} {\bibfnamefont {K.~S.}\ \bibnamefont {Murray}}, \ and\ \bibinfo
  {author} {\bibfnamefont {J.~D.}\ \bibnamefont {Cashion}},\ }\href@noop {}
  {\bibfield  {journal} {\bibinfo  {journal} {Science}\ }\textbf {\bibinfo
  {volume} {298}},\ \bibinfo {pages} {1762} (\bibinfo {year}
  {2002})}\BibitemShut {NoStop}%
\bibitem [{\citenamefont {G{\"u}tlich}\ and\ \citenamefont
  {Hauser}(1990)}]{gutlich1990thermal}%
  \BibitemOpen
  \bibfield  {author} {\bibinfo {author} {\bibfnamefont {P.}~\bibnamefont
  {G{\"u}tlich}}\ and\ \bibinfo {author} {\bibfnamefont {A.}~\bibnamefont
  {Hauser}},\ }\href@noop {} {\bibfield  {journal} {\bibinfo  {journal}
  {Coordination chemistry reviews}\ }\textbf {\bibinfo {volume} {97}},\
  \bibinfo {pages} {1} (\bibinfo {year} {1990})}\BibitemShut {NoStop}%
\bibitem [{\citenamefont {Chernyshov}\ \emph {et~al.}(2007)\citenamefont
  {Chernyshov}, \citenamefont {Klinduhov}, \citenamefont {T{\"o}rnroos},
  \citenamefont {Hostettler}, \citenamefont {Vangdal},\ and\ \citenamefont
  {B{\"u}rgi}}]{chernyshov2007coupling}%
  \BibitemOpen
  \bibfield  {author} {\bibinfo {author} {\bibfnamefont {D.}~\bibnamefont
  {Chernyshov}}, \bibinfo {author} {\bibfnamefont {N.}~\bibnamefont
  {Klinduhov}}, \bibinfo {author} {\bibfnamefont {K.~W.}\ \bibnamefont
  {T{\"o}rnroos}}, \bibinfo {author} {\bibfnamefont {M.}~\bibnamefont
  {Hostettler}}, \bibinfo {author} {\bibfnamefont {B.}~\bibnamefont {Vangdal}},
  \ and\ \bibinfo {author} {\bibfnamefont {H.-B.}\ \bibnamefont {B{\"u}rgi}},\
  }\href@noop {} {\bibfield  {journal} {\bibinfo  {journal} {Physical Review
  B}\ }\textbf {\bibinfo {volume} {76}},\ \bibinfo {pages} {014406} (\bibinfo
  {year} {2007})}\BibitemShut {NoStop}%
\bibitem [{\citenamefont {Konishi}\ \emph {et~al.}(2008)\citenamefont
  {Konishi}, \citenamefont {Tokoro}, \citenamefont {Nishino},\ and\
  \citenamefont {Miyashita}}]{konishi2008monte}%
  \BibitemOpen
  \bibfield  {author} {\bibinfo {author} {\bibfnamefont {Y.}~\bibnamefont
  {Konishi}}, \bibinfo {author} {\bibfnamefont {H.}~\bibnamefont {Tokoro}},
  \bibinfo {author} {\bibfnamefont {M.}~\bibnamefont {Nishino}}, \ and\
  \bibinfo {author} {\bibfnamefont {S.}~\bibnamefont {Miyashita}},\ }\href@noop
  {} {\bibfield  {journal} {\bibinfo  {journal} {Physical review letters}\
  }\textbf {\bibinfo {volume} {100}},\ \bibinfo {pages} {067206} (\bibinfo
  {year} {2008})}\BibitemShut {NoStop}%
\bibitem [{\citenamefont {Ohkoshi}\ \emph {et~al.}(2011)\citenamefont
  {Ohkoshi}, \citenamefont {Imoto}, \citenamefont {Tsunobuchi}, \citenamefont
  {Takano},\ and\ \citenamefont {Tokoro}}]{ohkoshi2011light}%
  \BibitemOpen
  \bibfield  {author} {\bibinfo {author} {\bibfnamefont {S.-i.}\ \bibnamefont
  {Ohkoshi}}, \bibinfo {author} {\bibfnamefont {K.}~\bibnamefont {Imoto}},
  \bibinfo {author} {\bibfnamefont {Y.}~\bibnamefont {Tsunobuchi}}, \bibinfo
  {author} {\bibfnamefont {S.}~\bibnamefont {Takano}}, \ and\ \bibinfo {author}
  {\bibfnamefont {H.}~\bibnamefont {Tokoro}},\ }\href@noop {} {\bibfield
  {journal} {\bibinfo  {journal} {Nature chemistry}\ }\textbf {\bibinfo
  {volume} {3}},\ \bibinfo {pages} {564} (\bibinfo {year} {2011})}\BibitemShut
  {NoStop}%
\bibitem [{\citenamefont {Pinkowicz}\ \emph {et~al.}(2015)\citenamefont
  {Pinkowicz}, \citenamefont {Rams}, \citenamefont {Mišek}, \citenamefont
  {Kamenev}, \citenamefont {Tomkowiak}, \citenamefont {Katrusiak},\ and\
  \citenamefont {Sieklucka}}]{pinkowicz2015enforcing}%
  \BibitemOpen
  \bibfield  {author} {\bibinfo {author} {\bibfnamefont {D.}~\bibnamefont
  {Pinkowicz}}, \bibinfo {author} {\bibfnamefont {M.}~\bibnamefont {Rams}},
  \bibinfo {author} {\bibfnamefont {M.}~\bibnamefont {Mišek}}, \bibinfo
  {author} {\bibfnamefont {K.~V.}\ \bibnamefont {Kamenev}}, \bibinfo {author}
  {\bibfnamefont {H.}~\bibnamefont {Tomkowiak}}, \bibinfo {author}
  {\bibfnamefont {A.}~\bibnamefont {Katrusiak}}, \ and\ \bibinfo {author}
  {\bibfnamefont {B.}~\bibnamefont {Sieklucka}},\ }\href@noop {} {\bibfield
  {journal} {\bibinfo  {journal} {Journal of the American Chemical Society}\
  }\textbf {\bibinfo {volume} {137}},\ \bibinfo {pages} {8795} (\bibinfo {year}
  {2015})}\BibitemShut {NoStop}%
\bibitem [{\citenamefont {Chernyshov}\ \emph {et~al.}(2004)\citenamefont
  {Chernyshov}, \citenamefont {B{\"u}rgi}, \citenamefont {Hostettler},\ and\
  \citenamefont {T{\"o}rnroos}}]{chernyshov2004landau}%
  \BibitemOpen
  \bibfield  {author} {\bibinfo {author} {\bibfnamefont {D.}~\bibnamefont
  {Chernyshov}}, \bibinfo {author} {\bibfnamefont {H.-B.}\ \bibnamefont
  {B{\"u}rgi}}, \bibinfo {author} {\bibfnamefont {M.}~\bibnamefont
  {Hostettler}}, \ and\ \bibinfo {author} {\bibfnamefont {K.~W.}\ \bibnamefont
  {T{\"o}rnroos}},\ }\href@noop {} {\bibfield  {journal} {\bibinfo  {journal}
  {Physical Review B}\ }\textbf {\bibinfo {volume} {70}},\ \bibinfo {pages}
  {094116} (\bibinfo {year} {2004})}\BibitemShut {NoStop}%
\bibitem [{\citenamefont {Nesterov}\ \emph {et~al.}(2017)\citenamefont
  {Nesterov}, \citenamefont {Orlov}, \citenamefont {Ovchinnikov},\ and\
  \citenamefont {Nikolaev}}]{nesterov2017cooperative}%
  \BibitemOpen
  \bibfield  {author} {\bibinfo {author} {\bibfnamefont {A.~I.}\ \bibnamefont
  {Nesterov}}, \bibinfo {author} {\bibfnamefont {Y.~S.}\ \bibnamefont {Orlov}},
  \bibinfo {author} {\bibfnamefont {S.~G.}\ \bibnamefont {Ovchinnikov}}, \ and\
  \bibinfo {author} {\bibfnamefont {S.~V.}\ \bibnamefont {Nikolaev}},\
  }\href@noop {} {\bibfield  {journal} {\bibinfo  {journal} {Physical Review
  B}\ }\textbf {\bibinfo {volume} {96}},\ \bibinfo {pages} {134103} (\bibinfo
  {year} {2017})}\BibitemShut {NoStop}%
\bibitem [{\citenamefont {Rodr{\'\i}guez-Castellanos}, \citenamefont
  {Plasencia-Montesinos},\ and\ \citenamefont
  {Reguera-Ruiz}(2018)}]{rodriguez2018spin}%
  \BibitemOpen
  \bibfield  {author} {\bibinfo {author} {\bibfnamefont {C.}~\bibnamefont
  {Rodr{\'\i}guez-Castellanos}}, \bibinfo {author} {\bibfnamefont
  {Y.}~\bibnamefont {Plasencia-Montesinos}}, \ and\ \bibinfo {author}
  {\bibfnamefont {E.}~\bibnamefont {Reguera-Ruiz}},\ }\href@noop {} {\bibfield
  {journal} {\bibinfo  {journal} {Revista Cubana de F{\'\i}sica}\ }\textbf
  {\bibinfo {volume} {35}},\ \bibinfo {pages} {91} (\bibinfo {year}
  {2018})}\BibitemShut {NoStop}%
\bibitem [{\citenamefont {Palii}\ \emph {et~al.}(2015)\citenamefont {Palii},
  \citenamefont {Ostrovsky}, \citenamefont {Reu}, \citenamefont {Tsukerblat},
  \citenamefont {Decurtins}, \citenamefont {Liu},\ and\ \citenamefont
  {Klokishner}}]{palii2015microscopic}%
  \BibitemOpen
  \bibfield  {author} {\bibinfo {author} {\bibfnamefont {A.}~\bibnamefont
  {Palii}}, \bibinfo {author} {\bibfnamefont {S.}~\bibnamefont {Ostrovsky}},
  \bibinfo {author} {\bibfnamefont {O.}~\bibnamefont {Reu}}, \bibinfo {author}
  {\bibfnamefont {B.}~\bibnamefont {Tsukerblat}}, \bibinfo {author}
  {\bibfnamefont {S.}~\bibnamefont {Decurtins}}, \bibinfo {author}
  {\bibfnamefont {S.-X.}\ \bibnamefont {Liu}}, \ and\ \bibinfo {author}
  {\bibfnamefont {S.}~\bibnamefont {Klokishner}},\ }\href@noop {} {\bibfield
  {journal} {\bibinfo  {journal} {The Journal of chemical physics}\ }\textbf
  {\bibinfo {volume} {143}},\ \bibinfo {pages} {084502} (\bibinfo {year}
  {2015})}\BibitemShut {NoStop}%
\bibitem [{\citenamefont {Cardy}(1996)}]{cardy1996scaling}%
  \BibitemOpen
  \bibfield  {author} {\bibinfo {author} {\bibfnamefont {J.}~\bibnamefont
  {Cardy}},\ }\href@noop {} {\emph {\bibinfo {title} {Scaling and
  renormalization in statistical physics}}},\ Vol.~\bibinfo {volume} {5}\
  (\bibinfo  {publisher} {Cambridge university press},\ \bibinfo {year}
  {1996})\BibitemShut {NoStop}%
\bibitem [{\citenamefont {Kardar}(2007)}]{kardar2007statistical}%
  \BibitemOpen
  \bibfield  {author} {\bibinfo {author} {\bibfnamefont {M.}~\bibnamefont
  {Kardar}},\ }\href@noop {} {\emph {\bibinfo {title} {Statistical physics of
  fields}}}\ (\bibinfo  {publisher} {Cambridge University Press},\ \bibinfo
  {year} {2007})\BibitemShut {NoStop}%
\bibitem [{\citenamefont {Newman}\ and\ \citenamefont
  {Barkema}(1999)}]{Newman99}%
  \BibitemOpen
  \bibfield  {author} {\bibinfo {author} {\bibfnamefont {M.}~\bibnamefont
  {Newman}}\ and\ \bibinfo {author} {\bibfnamefont {G.}~\bibnamefont
  {Barkema}},\ }\href@noop {} {\emph {\bibinfo {title} {Monte carlo methods in
  statistical physics chapter 1-4}}}\ (\bibinfo  {publisher} {Oxford University
  Press: New York, USA},\ \bibinfo {year} {1999})\BibitemShut {NoStop}%
\bibitem [{\citenamefont {Murray}\ and\ \citenamefont
  {Kepert}(2004)}]{murray2004cooperativity}%
  \BibitemOpen
  \bibfield  {author} {\bibinfo {author} {\bibfnamefont {K.~S.}\ \bibnamefont
  {Murray}}\ and\ \bibinfo {author} {\bibfnamefont {C.~J.}\ \bibnamefont
  {Kepert}},\ }in\ \href@noop {} {\emph {\bibinfo {booktitle} {Spin Crossover
  in Transition Metal Compounds I}}}\ (\bibinfo  {publisher} {Springer},\
  \bibinfo {year} {2004})\ pp.\ \bibinfo {pages} {195--228}\BibitemShut
  {NoStop}%
\bibitem [{\citenamefont {Chadi}\ and\ \citenamefont
  {Cohen}(1973)}]{chadi1973special}%
  \BibitemOpen
  \bibfield  {author} {\bibinfo {author} {\bibfnamefont {D.}~\bibnamefont
  {Chadi}}\ and\ \bibinfo {author} {\bibfnamefont {M.~L.}\ \bibnamefont
  {Cohen}},\ }\href@noop {} {\bibfield  {journal} {\bibinfo  {journal}
  {Physical Review B}\ }\textbf {\bibinfo {volume} {8}},\ \bibinfo {pages}
  {5747} (\bibinfo {year} {1973})}\BibitemShut {NoStop}%
\end{thebibliography}%

\end{document}